\documentclass[twocolumn]{aastex631}
\usepackage{multirow}

\received{}
\revised{}
\accepted{}

\submitjournal{The Astrophysical Journal}

\shorttitle{Particle Acceleration around Galaxy Clusters}
\shortauthors{Ha et al.}

\begin{document}

\title{Cosmic Ray Acceleration and Nonthermal Radiation at Accretion Shocks in the Outer Regions of Galaxy Clusters}

\author[0000-0001-7670-4897]{Ji-Hoon Ha}
\affil{Department of Physics, College of Natural Sciences UNIST, Ulsan 44919, Korea}
\author[0000-0002-5455-2957]{Dongsu Ryu}
\affil{Department of Physics, College of Natural Sciences UNIST, Ulsan 44919, Korea}
\author[0000-0002-4674-5687]{Hyesung Kang}
\affiliation{Department of Earth Sciences, Pusan National University, Busan 46241, Korea}
\correspondingauthor{Hyesung Kang}\email{hskang@pusan.ac.kr}

\begin{abstract}

Cosmology models predict that external accretion shocks form in the outer region of galaxy clusters due to supersonic gas infall from filaments and voids in the cosmic web. 
They are characterized by high sonic and Alfv\'enic Mach numbers, $M_s\sim10-10^2$ and $M_A\sim10^2-10^3$, and propagate into weakly magnetized plasmas of $\beta\equiv P_g/P_B\gtrsim10^2$.
Although strong accretion shocks are expected to be efficient accelerators of cosmic rays (CRs), nonthermal signatures of shock-accelerated CRs around clusters have not been confirmed, and detailed acceleration physics at such shocks has yet to be understood.
In this study, we first establish through two-dimensional particle-in-cell simulations that at strong high-$\beta$ shocks electrons can be pre-energized via stochastic Fermi acceleration owing to the ion-Weibel instability in the shock transition region, possibly followed by injection into diffusive shock acceleration.
Hence, we propose that the models derived from conventional thermal leakage injection may be employed for the acceleration of electrons and ions at accretion shocks as well.
Applying these analytic models to numerical shock zones identified in structure formation simulations, we estimate nonthermal radiation, such as synchrotron and inverse-Compton (IC) emission due to CR electrons, and $\pi^0$-decay $\gamma$-rays due to CR protons, around simulated clusters.
Our models with the injection parameter, $Q\approx3.5-3.8$, predict synthetic synchrotron maps, which seem consistent with recent radio observations of the Coma cluster. 
However, the detection of nonthermal IC X-rays and $\gamma$-rays from accretion shocks would be quite challenging.
We suggest that the proposed analytic models may be adopted as generic recipes for CR production at cosmological shocks.

\end{abstract}


\section{Introduction} 
\label{s1}

\begin{deluxetable*}{c|c|c|c}[t]
\tablecaption{Properties and Acceleration Physics of Cosmological Shocks \label{t1}}
\tabletypesize{\scriptsize}
\tablecolumns{12}
\tablenum{1}
\tablewidth{0pt}
\tablehead{
\colhead{} &
\multicolumn{2}{|c|}{Weak Internal Shocks} &
\colhead{External Accretion Shocks}
}
\startdata
shock location$^{(a)}$ & \multicolumn{2}{|c|}{$r_{\rm sh} \lesssim r_{\rm vir}$} & $r_{\rm sh} \gtrsim r_{\rm vir}$\\
\hline
sonic Mach number & \multicolumn{2}{|c|}{$ M_s \sim 2-4$} & $ M_s \sim 5-10^2$\\
\hline
shock speed& \multicolumn{2}{|c|}{$ u_s \sim 500-3\times 10^3~{\rm km~s^{-1}}$} & $u_s \sim 20-1.5\times 10^3~{\rm km~s^{-1}}$\\
\hline
magnetization parameter$^{(b)}$ & \multicolumn{2}{|c|}{$\sigma \sim 8\times10^{-4}-6 \times10^{-3}$} & $\sigma \sim 10^{-7}-10^{-4}$\\
\hline
plasma $\beta = P_g/P_B$& \multicolumn{2}{|c|}{$\beta \sim 50 - 10^2$} & $\beta \gtrsim 10^2$\\
\hline
obliquity of magnetic field, $\mathbf{B}_0$  & $Q_{\perp}$ & $Q_{\parallel}$ & weak magnetic fields\\
\hline
\multirow{3}*{main microinstabilities}      & electron firehose &ion/ion beam & ion Weibel \\
                                                     & Alfv\'{e}n ion cyclotron $\&$ ion-mirror & resonant streaming & electron whistler \\
                                                     & electron whistler $\&$ electron-mirror & nonresonant Bell &  \\
\hline
accelerated species & mainly electrons & mainly protons & both protons \& electrons \\
\hline
\multirow{2}*{main preacceleration to $p_{\rm inj}$} & Shock Drift Acceleration (SDA) & Fermi I-like & electron Fermi II \\
                                                                        & Stochastic SDA (SSDA)&  & proton (not observed) \\
\hline
acceleration beyond $p_{\rm inj}$ & DSA & DSA & DSA\\
\hline
\multirow{2}*{references} & \citet{guo2014a,kang2019} & \citet{ha2018b} & this work \\
                          & \citet{kobzar2021, ha2021} &  &  \\
\hline
\enddata 
\tablenotetext{a}{$r_{\rm sh}$ and $r_{\rm vir}$ stand for the shock location from the cluster center and the virial radius of clusters, respectively.}
\tablenotetext{b}{$\sigma=M_A^{-2}= \mathcal{E}_B/\mathcal{E}_{\rm sh}$, where $M_A$ is the Alfv\'enic Mach number.}

\end{deluxetable*}

In the current $\Lambda$CDM paradigm, the large-scale structure of the universe forms through hierarchical clustering of substructures. 
The supersonic flow motions associated with infall of baryonic gas toward sheets and filaments, as well as cluster mergers, naturally induce shocks in the cosmic web \citep[e.g.,][]{ryu2003}. 
The properties and roles of such cosmological shocks have been extensively studied using cosmological hydrodynamic simulations \citep[e.g.,][]{miniati2000, pfrommer2006, kang2007, hoeft2008, skillman2008, vazza2009, hong2014, schaal2015, ha2018a}. 
In particular, shocks associated with galaxy clusters can be classified mainly into two categories (see Table \ref{t1}). 
{\it Internal shocks} appear in the hot intracluster medium (ICM with $T \sim 10^7 - 10^8$ K) within the virial radius (i.e, $r_{\rm sh} < r_{\rm vir}$) due to continuous mergers of substructures and turbulent flow motions, and they have the sonic Mach number, $M_s \sim 2 - 4$, and the plasma beta, $\beta\equiv P_g/P_B\sim 50-100$. Here, $P_g$ and $P_B$ are the gas and magnetic pressures, respectively.
On the other hand, external {\it accretion shocks} form in the outer region well outside of the virial radius (i.e., $r_{\rm sh} > r_{\rm vir}$), where the cold gas in void regions ($T \sim 10^4$ K) and the warm-hot intergalactic medium (WHIM) in filaments ($T \sim 10^5 - 10^7$ K) accrete onto clusters, and so the sonic and Alfv\'enic Mach numbers tend to be much higher, $M_s \sim 10 - 10^2$ and $M_A\sim 10^2-10^3$. 

As for typical astrophysical shocks, these cosmological shocks are collisionless, and hence are expected to accelerate cosmic ray (CR) protons and electrons to very high energies via diffusive shock acceleration (DSA, also known as Fermi I acceleration). 
Its central idea stands on a rather solid ground: CR particles can gain energy by crossing multiple times the shock front through scattering off underlying MHD waves \citep[e.g.,][]{bell1978, blandford1978, drury1983}.
Thus, this mechanism requires certain prerequisite processes, such as the pre-energization of suprathermal particles, the self-excitation of kinetic/MHD waves by a variety of microinstabilities, and/or the presence of preexisting magnetic turbulence.

Because the width of the shock transition zone is roughly a few times the gyroradius of thermal protons, both protons and electrons need to be energized to the so-called injection momentum, $p_{\rm inj} \sim 3.5 p_{\rm th,p}$, in order to participate fully in the DSA process \citep[e.g.,][]{kang2010,caprioli2014a, park2015,ha2018b}. 
Here, $p_{\rm th,p}= \sqrt{2m_p k_B T_2}$ is the proton thermal momentum in the postshock gas, $T_2$ is the post temperature, $m_p$ is the proton mass, and $k_B$ is the Boltzmann constant.
This processes is referred to as particle {\it preacceleration} or {\it injection}.
In particular, electron injection is a much more challenging problem,
because the electron thermal momentum, $p_{\rm th,e}=\sqrt{m_e/m_p}~p_{\rm th,p}$, is much smaller than $p_{\rm inj}$
(here, the proton to electron mass ratio is $m_p/m_e = 1836$).

Evidently, the magnetic field and its fluctuations play crucial roles in the acceleration of nonthermal particles as well as the formation of collisionless shocks.
Such kinetic processes are known to depend on the configuration of the background magnetic fields, $\mathbf{B}_0$, as specified in Table \ref{t1}. 
Specifically, protons are accelerated mainly at quasi-parallel ($Q_{\parallel}$) shocks with $\theta_{\rm Bn} \lesssim 45^{\circ}$ \citep[e.g.,][]{ha2018b}, while electrons are preferentially accelerated at quasi-perpendicular ($Q_{\perp}$) shocks with $\theta_{\rm Bn} \gtrsim 45^{\circ}$ \citep[e.g.,][]{guo2014a,guo2014b, kang2019}, where $\theta_{\rm Bn}$ is the obliquity angle between the shock normal and $\mathbf{B}_0$.
In general, the physics of collisionless shocks depends on various shock parameters such as $M_s$, $\beta$, and $\theta_{\rm Bn}$ \citep[e.g.,][]{balogh2013}.

In the case of nonrelativistic, high-Mach-number shocks ($M_s\sim M_A>10$) in $\beta\lesssim 1$ plasmas, such as supernova remnants and planetary bow shocks,
the injection and acceleration of CR protons (CRp) and CR electrons (CRe) have been relatively well established through many studies using hybrid and particle-in-cell (PIC) simulations \citep[e.g.,][]{caprioli2014a,caprioli2014b,amano2009,riquelme2011,park2015,marcowith2016}.
In the $Q_{\perp}$ magnetic field configuration, for instance, particles may undergo the gradient-drift along the shock surface due to the compressed magnetic field, and gain energy from the shock motional electric field, $-(\mathbf{u_{\rm sh}}/c)\times \mathbf{B}_0$.
This is known as shock drift acceleration (SDA). 

In particular, in $Q_{\perp}$ shocks with high Alfv\'enic Mach numbers ($M_A\gtrsim 1.2 (m_p/m_e)^{2/3}$) or low magnetization parameter ($\sigma=M_A^{-2}$), it was shown that
electrons are preaccelerated through the following processes: 
(1) the shock surfing acceleration (SSA) due to electron trapping by electrostatic Buneman waves and the interaction with the motional electric field at the leading edge of the shock foot, (2) the stochastic Fermi II acceleration due to multiple wave-particle interactions mediated by Weibel-induced filaments at the shock ramp, (3) the stochastic shock drift acceleration (SSDA) due to electron confinement by multi-scale waves in the shock overshoot, and 
(4) magnetic reconnection in Weibel-induced filaments \citep[e.g.,][]{matsumoto2012,matsumoto2015, matsumoto2017, bohdan2017, bohdan2019a,bohdan2019b, bohdan2020a,katou2019,amano2022}. 

Electron preacceleration at {\it weak internal shocks} ($M_s\lesssim 4$) in the high-$\beta$ ICM 
was examined by a number of earlier studies \citep[e.g.,][]{guo2014a,guo2014b,ha2018b, kang2019, niemiec2019, ha2021,kobzar2021}. 
With weak background magnetic fields relevant for the ICM ($\beta \sim 50 - 100$), protons and electrons of the incoming plasma could be reflected by the shock electrostatic potential drop and/or the magnetic mirror force, undergoing SDA at the shock ramp. 
Mainly, two types of electron preacceleration mechanisms have been proposed:
(1) the so-called {\it Fermi-like SDA} due to multiple cycles of SDA and diffusive scattering of electrons between the shock ramp and upstream self-generated waves \citep{guo2014a, kang2019},
and (2) the SSDA due to the extended gradient-drift of electrons, while being confined in the shock overshoot \citep{niemiec2019, ha2021,kobzar2021}. 
Reflected particles result in the ion and electron temperature anisotropies, 
which in turn excite multi-scale plasma waves via various microinstabilities: e.g.,
the electron firehose instability in the upstream region \citep{guo2014b,kim2020},
the Alfv\'{e}n ion cyclotron and ion-mirror instabilities,
and the electron whistler and electron-mirror instabilities in the shock overshoot \citep[e.g.,][]{katou2019,kim2021}.

Although cluster accretion shocks have been proposed as efficient particle accelerators in some previous studies \citep[e.g.,][]{kang1996, ryu2003, kang2013, hong2014, vazza2016}, the kinetic plasma physics at strong shocks in high $\beta\gtrsim 10^2$ plasmas has yet to be explored in detail. 
Since the initial background magnetic field in the upstream void and filament regions is expected to be very weak, 
the formation of accretion shocks could be mediated by self-excited magnetic field fluctuations induced by the so-called Weibel instability.
As shown in Table \ref{t1}, accretion shocks are characterized by high Mach numbers, $M_s\sim 10-10^2$ and $M_A\sim 10^2-10^3$, and high plasma beta, $\beta\sim 10^2-10^3$. 
The corresponding magnetization parameter is quite low, $\sigma = \mathcal{E}_B/ \mathcal{E}_{\rm sh} \sim 10^{-7}-10^{-4}$, where $\mathcal{E}_B = B_0^2/8\pi$ is the background magnetic energy density and $\mathcal{E}_{\rm sh}=(1/2)n_1 m_p u_s^2$ is the shock kinetic energy density.

The Weibel instability is known to be excited by the anisotropy in the proton velocity space due to the shock-reflected protons at high-$M_A$ (or low-$\sigma$) shocks.
In general, it can be classified into two categories: (1) the original Weibel instability due to the temperature anisotropy ($T_{\perp} > T_{\parallel}$) with respect to the background magnetic field \citep{Weibel1959}, and (2) the beam-Weibel instability, also known as the filamentation instability, due to the counter-streaming beams \citep{fried1959}. 
Although the relevant instability in the shock transition region is the filamentation instability due to the interactions between the shock-reflected and incoming ions, for the sake of simplicity, we use the term ``Weibel instability" throughout this paper. 
This instability is known to be responsible for the self-generation and amplification of magnetic fields in initially unmagnetized or weakly magnetized plasmas \citep[e.g.,][]{schlickeiser2003}.
  
The formation of Weibel-mediated shocks, magnetic field amplification, and ensuing particle acceleration have been investigated through PIC simulations for both relativistic shocks \citep[e.g.,][]{ spitkovsky2008,sironi2009,sironi2013} 
and nonrelativistic shocks \citep[e.g.,][]{kato2008, kato2010, bohdan2021}, as well as
in laboratory laser experiments \citep[e.g.,][]{fiuza2012,huntington2015}. 
In particular, it has been shown that in the nonrelativistic, {\it unmagnetized} flows the Weibel instability can amplify the magnetic field energy, up to $\sim 1 \%$ of the bulk kinetic energy \citep[e.g.,][]{kato2008, huntington2015}.
In addition, recent PIC simulations indicate the evidence of full DSA at Weibel-mediated shocks \citep[][]{fiuza2020, arbutina2021}.
These previous studies focused on high-$M_A$ shocks in relatively low $\beta\lesssim 1$ plasmas.
In the current study, by contrast, we examine the shock formation and electron preacceleration in the regime of $M_A\gg M_s\gg 1$ with $\beta\sim 10^2-10^3$.

Interactions between CR particles and the ambient medium, radiation, and magnetic fields generate nonthermal emission in a wide range of wavelengths. 
For instance, double radio relics detected in merging clusters could be understood as radio synchrotron emission from CRe that are accelerated at merger-driven, internal shocks in the ICM \citep[e.g.,][]{vanweeren2010,vanweeren2011,vanweeren2019}. 
Although the presence of CRp in the ICM has yet to be established through a positive detection of cluster-wide $\gamma$-rays from $\pi^0$-decay after inelastic CRp-p collisions \citep[e.g.,][]{pinzke2010,vazza2016,ackermann2016,ryu2019, wittor2020,adam2021,mirakhor2022}, the relative abundance of CRp to CRe is known to be about $K_{p/e}\sim 100$ in the Galaxy; hence, CRp are expected to be injected and accelerated with a higher efficiency than CRe at ICM shocks. 
 
Nonthermal emission in the outer region of galaxy clusters beyond the virial radius 
has long been postulated as direct evidence for CR acceleration at strong accretion shocks. 
In particular, $\gamma$-ray \citep[e.g.,][]{loeb2000, totani2000, miniati2003,scharf2002, pinzke2010,keshet2003} and hard X-ray (HXR) radiation \citep[e.g.,][]{ensslin1999, kushnir2010} could possibly be produced by the inverse-Compton (IC) scattering of cosmic microwave background (CMB) photons off CRe accelerated at accretion shocks. 
Radio synchrotron radiation could also originate from accretion shocks as well \citep[e.g.,][]{bonafede2022}. 
Although a few studies argue that the $\gamma$-ray ring and radio synchrotron emission might have been observed around the Coma cluster \citep[e.g.,][]{keshet2017,bonafede2022}, more observational evidence would be required to firmly support CRe acceleration at accretion shocks.
On the other hand, $\gamma$-ray emission from $\pi^0$-decay due to inelastic $p$-$p$ collision is expected to be produced mostly in the dense inner region, rather than outer region of clusters \citep[e.g.,][]{miniati2003,pinzke2010}.

In this paper, we examine nonthermal radiation from cluster accretions shocks by the following three steps.
(1) In Section \ref{s2}, we demonstrate that electrons could be preaccelerated and injected into DSA via the stochastic Fermi II process owing to the ion-beam-Weibel instability in the shock transition region of strong shocks with $M_A\sim 10^2-10^3$.
(2) In Section \ref{s3}, we presume that both protons and electrons are accelerated to form the canonical power-law spectra via the DSA process in the test-particle regime at both internal and accretion shocks. In fact, we follow the classical thermal leakage injection recipe for both CRp and CRe, in which the power-law spectra emerge directly from the postshock Maxwellian distributions \citep[e.g.,][]{kang2010, caprioli2014a, ryu2019, kang2020,arbutina2021}.
(3) In Section \ref{s4}, as a post-processing, we apply the analytic DSA model spectra for CRp and CRe to numerically identified shocks in structure formation simulations \citep[e.g.,][]{ryu2003,ha2018a}. We then estimate the nonthermal emission from the shock-accelerated CRp and CRe, such as electron synchrotron radiation, electron IC radiation in X-ray and $\gamma$-ray bands, and proton $\pi^0$ $\gamma$-rays. 
We then discuss the implications for multi-wavelength observations of nonthermal radiation in the outer region of galaxy clusters.
Finally, we give a brief summary of this work in Section \ref{s5}. 

\begin{deluxetable*}{ccccccccccccc}[t]
\tablecaption{Model Parameters for 2D PIC Simulations \label{t2}}
\tabletypesize{\scriptsize}
\tablecolumns{12}
\tablenum{2}
\tablewidth{0pt}
\tablehead{
\colhead{Model Name} &
\colhead{$M_{s}$} &
\colhead{$M_{A}$} &
\colhead{$u_0/c$} &
\colhead{$\beta$} &
\colhead{$T_0 [K]$} &
\colhead{$m_i/m_e$} &
\colhead{$B$-field} &
\colhead{$L_x[c/\omega_{\rm pe}]$}&
\colhead{$L_y[c/\omega_{\rm pe}]$}&
\colhead{$L_x[c/\omega_{\rm pi}]$}&
\colhead{$L_y[c/\omega_{\rm pi}]$}&
\colhead{$t_{\rm end} [\omega_{\rm pi}^{-1}]$}
}
\startdata
m100-Ly1.0$^{(a)}$ &  100 &  $2.9\times10^3$ &  0.018 & $10^3$ & $10^4$ & 100 & in-plane & 2000 & 300 & 200 & 30  & $8.0 \times 10^3$\\
\hline
m50-Ly0.5 &  100 &$2.9\times10^3$& 0.024 & $10^3$ & $10^4$ & 50 & in-plane & 1424 & 107 & 200 & 15 & $8.0 \times 10^3$\\
m50-Ly1.0 &  100 &$2.9\times10^3$& 0.024 & $10^3$ & $10^4$ & 50 & in-plane & 1424 & 213 & 200 & 30 & $8.0 \times 10^3$\\
m50-Ly1.5 &  100 &$2.9\times10^3$& 0.024 & $10^3$ & $10^4$ & 50 & in-plane & 1424 & 320 & 200 & 45 & $8.0 \times 10^3$\\
m25-Ly1.0 &  100 &$2.9\times10^3$& 0.036 & $10^3$ & $10^4$ & 25 & in-plane & 1000 & 150 & 200 & 30 & $8.0 \times 10^3$\\
\hline
m50-Ly1.0-M50 &  50 &$1.4\times10^3$& 0.012 & $10^3$ & $10^4$ & 50 & in-plane & 1424 & 213 & 200 & 30 & $8.0 \times 10^3$\\
m50-Ly1.0-M25 &  25 &$7.2\times10^2$& 0.006 & $10^3$ & $10^4$ & 50 & in-plane & 1424 & 213 & 200 & 30 & $8.0 \times 10^3$\\
\hline
m50-Ly1.0-Bz &  100 &$2.9\times10^3$& 0.024 & $10^3$ & $10^4$ & 50 & out-of-plane & 1424 & 213 & 200 & 30 & $8.0 \times 10^3$\\
\hline
m50-Ly1.0-M10-T6 &  10 &$0.9\times10^2$& 0.024 & $10^2$ & $10^6$ & 50 & in-plane & 1424 & 213 & 200 & 30 & $8.0 \times 10^3$\\
\enddata
\tablenotetext{a}{The fiducial model.}
\vspace{-0.2cm}
\end{deluxetable*}

\section{Electron Preacceleration at Strong High-$\beta$ Shocks}
\label{s2}

\subsection{Numerics for 2D PIC Simulations}
\label{s2.1}

Two-dimensional (2D) PIC simulations were performed using a parallel electromagnetic PIC code, TRISTAN-MP \citep[][]{buneman1993, spitkovsky2005}. 
All three components of particle velocities and electromagnetic fields are tracked, whereas particle positions are followed only in the $x$-$y$ simulation domain.
We employ an incoming ion-electron plasma flow with the velocity $-u_0 \mathbf{\hat{x}}$, which is stopped at the reflecting wall at the left side of the computational domain.
A collisionless shock is generated via interaction between the reflected and incoming flows and propagates to the $+x$-direction.
So the simulation frame corresponds to the downstream rest frame of the shock. 
A uniform background magnetic field of ${\mathbf{B_0} }= B_0 \mathbf{\hat{y}}$ (in-plane field case) or ${\mathbf{B_0} }= B_0 \mathbf{\hat{z}}$ (out-of-plane field case) is employed.

We adopt the plasma density of $n_0=10^{-4} {\rm cm^{-3}}$, relevant for the ICM in the cluster outskirt. 
Then, the temperature of the incoming plasma, $T_0 = T_{i} = T_{e}$, and the strength $B_0$ of the background field are chosen according to the plasma beta, $\beta = 16\pi n_0 k_B T/B_0^2$, as follows: (1) the gas accretion from void regions with $T_0 =10^4$ K and $\beta =10^3$, and (2) the gas accretion from filaments of galaxies with $T_0 =10^6$ K and $\beta =10^2$. 
For given $T_i$ and $B_0$, the sonic and Alfv\'{e}n Mach numbers are calculated as $M_{\rm s} = u_{\rm sh}/\sqrt{2\Gamma k_B T_i/m_i}$ and $M_{\rm A} = u_{\rm sh}/(B_0/\sqrt{4\pi n_0 m_i}) = M_{\rm s}\sqrt{\Gamma \beta/2}$, respectively. Here, the shock velocity in the upstream rest frame, $u_{\rm sh} = ru_0/(r-1)$, is estimated using the shock compression ratio, $r = (\Gamma+1)/(\Gamma - 1 + 2/M_{\rm s}^2)$, where $\Gamma = 5/3$ is the adiabatic index of the ICM gas. 
In the downstream rest frame, the shock moves with the speed, $u_{\rm sh}^{\rm sim} \approx u_{\rm sh} - u_0 = u_{\rm sh}/(r-1)$.
The flow speed $u_0$ is set to achieve
the sonic Mach number of simulated shocks, in the range of $M_{\rm s} = 10 - 10^2$, relevant for the accretion shocks.

The spatial resolution is $\Delta x = \Delta y = 0.1 c/\omega_{\rm pe}$, and the time step is $\Delta t = 0.045 \omega_{\rm pe}^{-1}$, where $\omega_{\rm pe} = \sqrt{4\pi n_0 e^2/m_e}$ is the electron plasma frequency. We note that $\Delta x$, $\Delta y$, and $\Delta t$ are expressed in units of the electron skin depth, $c/\omega_{\rm pe}$, and the electron plasma oscillation period, $\omega_{\rm pe}^{-1}$, respectively.  
On the other hand, the shock structure parameters, including the width of the shock transition region and the length of Weibel-induced filaments, and the evolution timescale of the shock are better described in terms of the ion skin depth, $c/\omega_{\rm pi} = \sqrt{m_i/m_e}c/\omega_{\rm pe}$, and the ion plasma oscillation period, $\omega_{\rm pi}^{-1} = \sqrt{m_i/m_e}\omega_{\rm pe}^{-1}$, respectively. 
For all the models, the simulations are carried out up to $t_{\rm end} \approx 8.0 \times 10^3 \omega_{\rm pi}^{-1}$.

Reduced mass ratios, $m_i/m_e = 25 - 100$, are employed due to the limitations of available computational resources.\footnote{We use the term ``ion” to represent positively charged particles with a range of mass ratios, $m_i/m_e$, in PIC simulations. Throughout the paper, on the other hand, the subscript $p$ is used to denote protons with real physical mass, $m_p$.} 
The box size along the $x$-direction is set to be $L_{\rm x} = 200 c/\omega_{\rm pi}$, which is long enough to accommodate the shock transition region with Weibel-induced filaments, $\sim (10-100) c/\omega_{\rm pi}$ \citep[e.g.,][]{kato2008}.   
The transverse size of the simulation domain is $L_{\rm y} = (15 - 45) c/\omega_{\rm pi}$, which covers the relevant scales of the filaments, i.e., the width, $l_{f,w}\sim c/\omega_{\rm pi}$, and the mean separation, $l_{f,s}\sim (1 - 10) c/\omega_{\rm pi}$ \citep[e.g.,][]{kato2008, bohdan2017}.

The model parameters of our simulations are summarized in Table \ref{t2}. 
The fiducial model, m100-Ly1.0, is specified by $M_s=10^2$, $m_i/m_e=100$, $L_y=30c/\omega_{\rm pi}$, and an in-plane background magnetic field. 
The name of other models is characterized with the value of $m_i/m_e$, the relative size of the simulation domain in the $y$-direction, and possibly a few other parameters.
In the m50Ly1.5 model, for instance, $m_i/m_e=50$ and $L_y$ is 1.5 times larger than that of the fiducial case.
Similarly, the m50-Ly1.0-Bz model considers an out-of-plane magnetic field, 
while the m50-Ly1.0-M10-T6 model considers a Mach 10 shock with $T_1=10^6$~K. 

\subsection{Microinstabilities}
\label{s2.2}

As mentioned in the introduction, the presence of magnetic fluctuations on relevant kinetic scales is essential for electron preacceleration and injection to DSA. 
In high-$M_A$, $Q_{\perp}$ shocks with $\beta\lesssim 1$, electron preacceleration is known to proceed through the SSA mediated by Buneman-induced waves in the leading edge of the shock foot, and then stochastic Fermi II acceleration mediated by Weibel-induced magnetic turbulence in the shock foot and ramp \citep[e.g.][]{matsumoto2017, bohdan2017, bohdan2019a, bohdan2019b}.

The Buneman instability can generate coherent electrostatic waves, if the relative drift velocity between shock-reflected ions and incoming electrons is larger than the electron thermal velocity, i.e., $M_A \gtrsim 0.4 \beta^{1/2} (m_i/m_e)^{1/2}$ \citep{matsumoto2012}.
The electrostatic field of the Buneman waves lies in the simulation plane and has a typical length scale of $\lambda \sim 2\pi u_0/\omega_{\rm pe}$ and a maximum amplitude of $E_{\rm es} \sim B_0$ \citep{amano2009,matsumoto2012}.
By contrast, the amplitude of the motional electric field along the $z$ axis is $E_{\rm mot}= (u_s/c) B_0 \ll E_{\rm es}$ for non-relativistic shocks.
So in earlier studies for $\beta\lesssim 1$ shocks \citep[e.g.,][]{bohdan2019a}, the initial electron energization was found to be driven by the work done by the electrostatic field of the Buneman waves.

\begin{figure}[t]
\vskip 0.0 cm
\hskip -0.0 cm
\centerline{\includegraphics[width=0.5\textwidth]{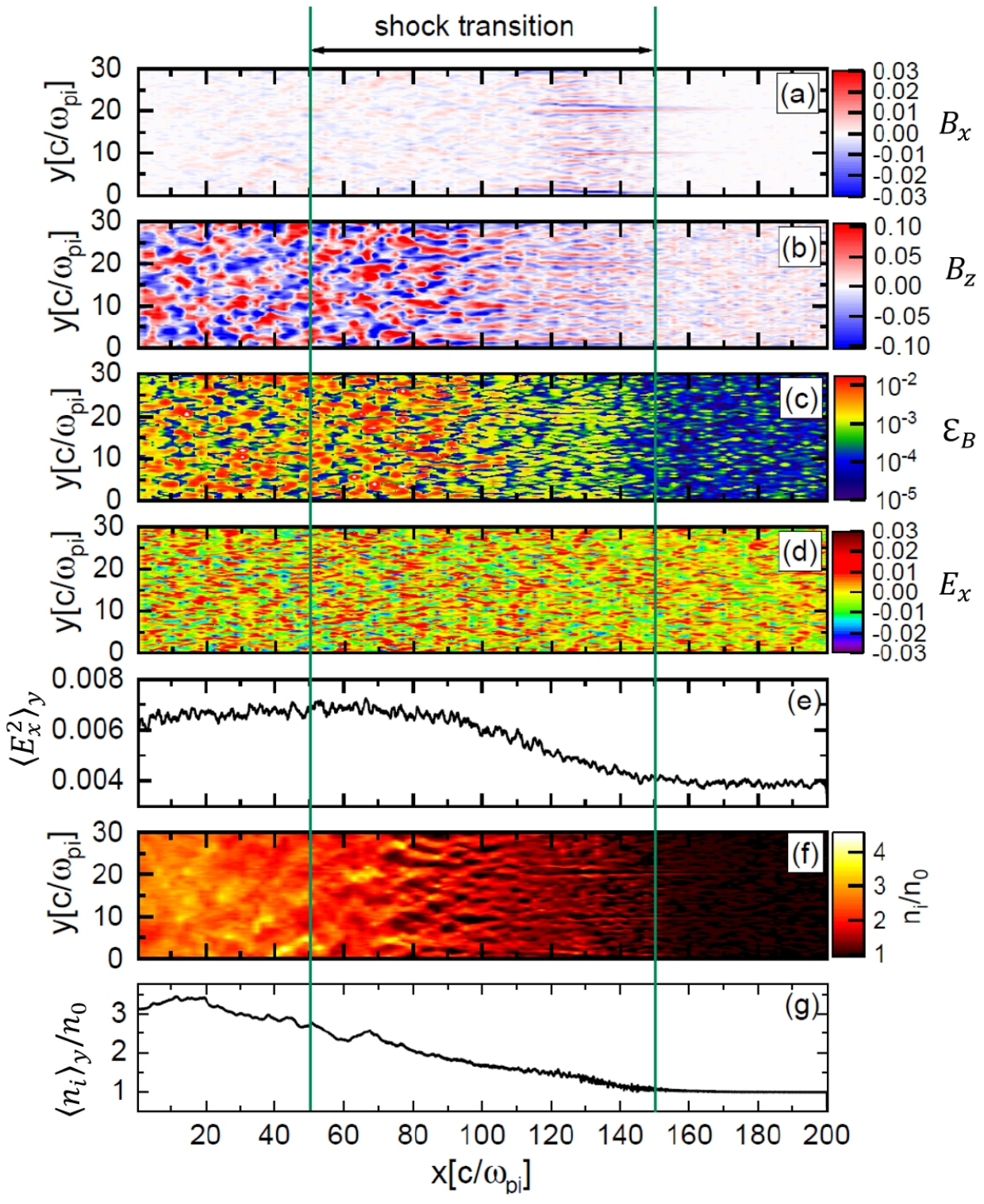}}
\vskip -0.0cm
\caption{Shock structures of the fiducial model, m100-Ly1.0, at $t_{\rm end} \sim 8.0 \times 10^3 \omega_{\rm pi}^{-1}$:
(a) magnetic field component $B_x$,
(b) magnetic field component $B_z$, 
(c) energy density of total magnetic field fluctuations $\mathcal{E}_B = \delta B^2/8\pi$, 
(d) electric field component $E_x$,
(e) $y$-averaged profile of $\langle E_x^2\rangle_y$,
(f) ion number density $n_i/n_0$, and 
(g) $y$-averaged profile of $\langle n_i \rangle_y /n_0$.
Here, the magnetic and electric field components are normalized by $\sqrt{8\pi \mathcal{E}_{\rm kin}}$, where $\mathcal{E}_{\rm kin} = (1/2)n_0 m_i u_0^2$ is the kinetic energy density of the upstream flow.
The shock transition region corresponds to $[x]\approx [50,150]c/\omega_{\rm pi}$.\label{f1}}
\end{figure}

However, the Buneman instability is expected to be suppressed by electron thermal motions in hot plasmas considered here.
In the fiducial model, for instance, the drift speed of the reflected ions relative to the incoming electrons, $\langle u_{\rm ref,ion}-u_{0e}\rangle \approx 0.048c$, is smaller than the thermal speed of electrons, $v_{\rm th,e}\approx 0.053c$, so the Buneman instability is likely to be stabilized in the high-$\beta$ regime under consideration \citep{matsumoto2012}.
Moreover, in the very low-$\sigma$ regime with weak $B_0$,
the Buneman waves on electron scales ($\sim c/\omega_{\rm pe}$) and the ensuing SSA process would be much less important,
because the Weibel magnetic fluctuations on ion scales ($\sim c/\omega_{\rm pi}$) dominate over any electron scale waves of relatively small amplitudes.
The 2D distribution of the electric field component $E_x$, shown in Figure \ref{f1}(d), does not exhibit prominent coherent electrostatic waves on electron scales.
Thus, we interpret that the Buneman instability is suppressed by electron thermal motions in the high-$\beta$ regime considered in here, as expected. 
We also note that the cross-shock potential is weakly induced in the almost unmagnetized shock models with $\beta=10^3$, so the $y$-averaged $\langle E_x^2 \rangle_y$ reveals only a gradual transition with a small amplitude in Figure \ref{f1}(e). 

\begin{figure}[t]
\vskip 0.0 cm
\hskip -0.2 cm
\centerline{\includegraphics[width=0.4\textwidth]{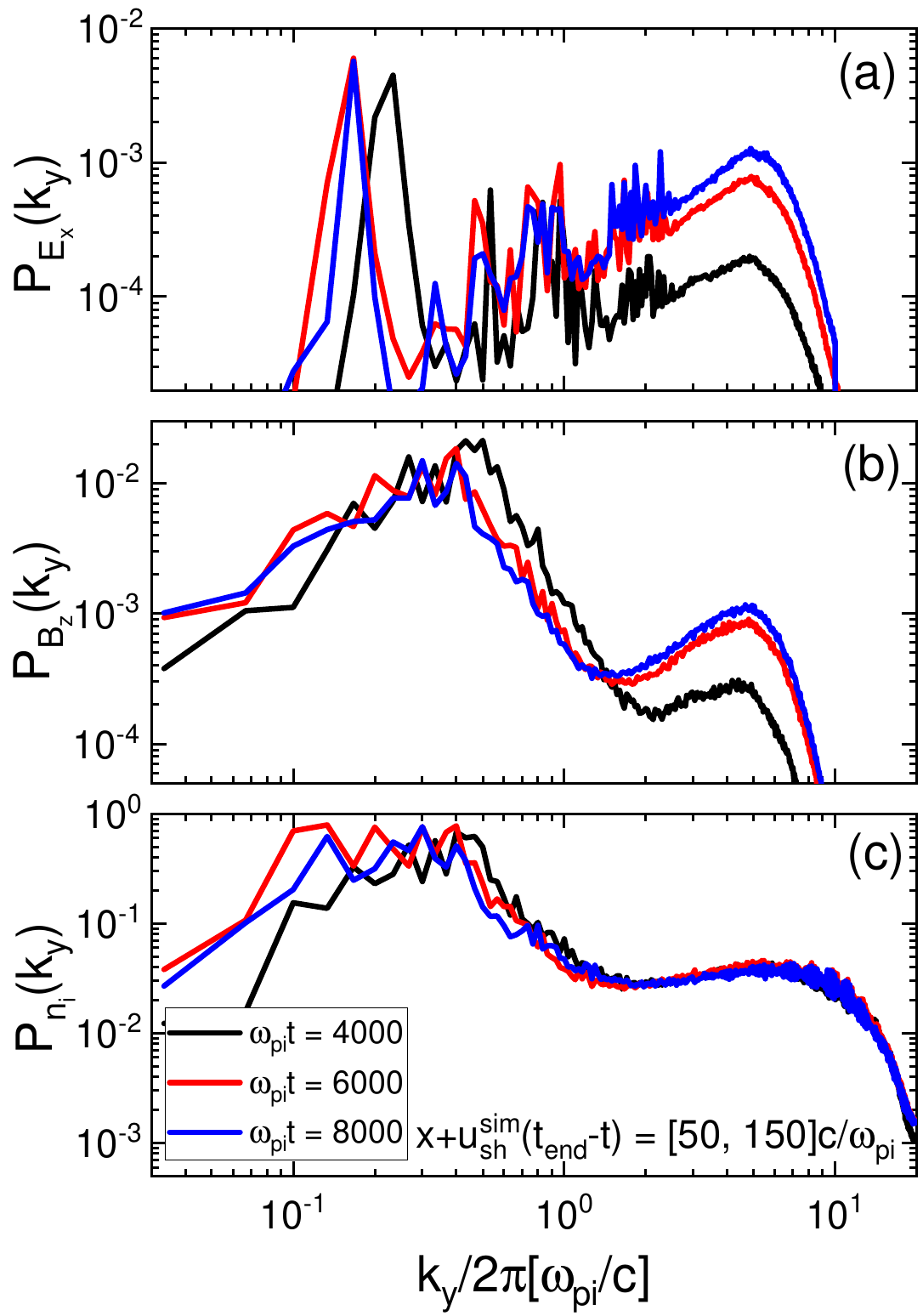}}
\vskip -0.2cm
\caption{Properties of the fluctuations in the shock transition region of the fiducial model. 
(a) Electric field power spectrum, $P_{E_x}(k_y) \propto (k_y/2\pi)\delta E_x^2$.
(b) Magnetic field power spectrum, $P_{B_z}(k_y) \propto (k_y/2\pi)\delta B_z^2$.
(c) Density power spectrum, $P_{n_i}(k_y) \propto (k_y/2\pi)\delta n_i^2$. 
Because the shock propagates with $u_{\rm sh}^{\rm sim}$ in the simulation frame, the shock transition region is defined by $[x+u_{\rm sh}^{\rm sim}(t_{\rm end}-t)] = [50 - 150]c/\omega_{\rm pi}$.
The simulation results during $t \sim (4.0 - 8.0) \times 10^3 \omega_{\rm pi}^{-1}$ are used.  \label{f2}}
\end{figure}

As demonstrated in a number of previous studies \citep[e.g.,][]{kato2010, bohdan2019b}, in high-$M_A$ shocks, filamentary magnetic structures are generated via the Weibel instability with the width $l_{f,w}\sim  c/\omega_{\rm pi}$ and the separation $l_{f,s}\sim (1 - 10) c/\omega_{\rm pi}$, which result in a broad shock transition region with the length scale of $\sim 100 c/\omega_{\rm pi}$.
Panels (a) and (b) of Figure \ref{f1} display the 2D spatial distributions of the two components of amplified magnetic fields, indicating $B_z\gg B_x$.
In the case of unmagnetized shocks, during the linear stage, the generated magnetic field is mostly transverse to the shock propagation direction and randomly oriented in the shock surface plane (the $y$-$z$ plane in our simulation setup), leading to the randomization of the particle velocity distribution through the Lorentz deflection.
Pronounced structures in the 2D spatial distributions of $B_z$, the energy density of the total magnetic field fluctuations $\mathcal{E}_B= \delta B^2/8\pi$, and the ion number density $n_i$, shown in panels (b), (c), and (f) of Figure \ref{f1}, clearly demonstrate the Weibel-induced filamentation in the shock transition. 

The magnetic field energy density increases up to the level of $\sim 1 \%$ of the bulk kinetic energy density, $\mathcal{E}_{\rm kin} = (1/2) n_0m_i u_0^2$. 
So the magnetic field is amplified up to $B \sim 0.1 \sqrt{8\pi \mathcal{E}_{\rm kin}}\sim 10^2 B_0$, which is much stronger than the initial background field. 
This is in a good agreement with previous PIC simulations and laboratory laser experiments \citep[e.g.,][]{kato2008,huntington2015}.
On the other hand, previous studies recognized that the Weibel amplified fields decrease on ion-gyro scales in the postshock region \citep[e.g.,][]{kato2008,bohdan2021}. Such $B$-field decay is observed in our simulations as well. Thus, in order to obtain amplified magnetic fields on macroscopic MHD scales, we have to rely on other processes, such as resonant CR streaming and nonresonant Bell instabilities \citep{bell1978,bell2004,caprioli2014b}, and turbulent dynamo amplification \citep{giacalone2007,ji2016}.

We also note that the perpendicular electron temperature with respect to the mean amplified field is larger than the parallel electron temperature ($T_{e\perp} > T_{e\parallel}$) in the shock transition, 
because Weibel-amplified magnetic fields, $\delta B_z$, are coherent on ion-scales of $\sim 1 - 10 c/\omega_{\rm pi}$ and electrons are energized via $E_{\perp}$ mainly in the simulation plane (see Figure \ref{f4}(g)-(h)).  
Consequently, this electron temperature anisotropy could excite the whistler instability and generate electron-scale waves.

In the shock transition region of $ [x + u_{\rm sh}^{\rm sim}(t_{\rm end} - t)] \approx [50 - 150] c/\omega_{\rm pi}$,
multi-scale fluctuations are induced as can be seen in Figure \ref{f1}. 
Figure \ref{f2} (a)-(c) shows the power spectra, $P_{E_x}(k_y) \propto k_y \delta E_x(k_y)^2$, $P_{B_z}(k_y) \propto k_y \delta B_z(k_y)^2$, and $P_{n_i}(k_y) \propto k_y \delta n_i(k_y)^2$,
in the shock transition during $t \sim (4000-8000) \omega_{\rm pi}^{-1}$. 
The three power spectra exhibit a peak in the range of $k_{y}/2\pi \sim (0.1 - 0.5) \omega_{\rm pi}/c$, which corresponds to the typical separation of Weibel filaments, $l_{f,s}\sim (2 - 10) c/\omega_{\rm pi}$.

In addition, both $P_{B_z}(k_y)$ and $ P_{E_x}(k_y)$ have the secondary peak at $k_y/2\pi \sim 5 \omega_{\rm pi}/c= 0.5\omega_{\rm pe}/c $ with a much smaller amplitude than that of the first peak.
This peak is likely to be related to electron-scale waves excited by the whistler instability due to the local electron temperature anisotropy.
Alternatively, it could result from interactions between weak Buneman-induced electrostatic waves and strong Weibel-induced electromagnetic waves.

\subsection{Electron Preacceleration}
\label{s2.3}

\begin{figure}[t]
\vskip 0.0 cm
\hskip -0.2 cm
\centerline{\includegraphics[width=0.45\textwidth]{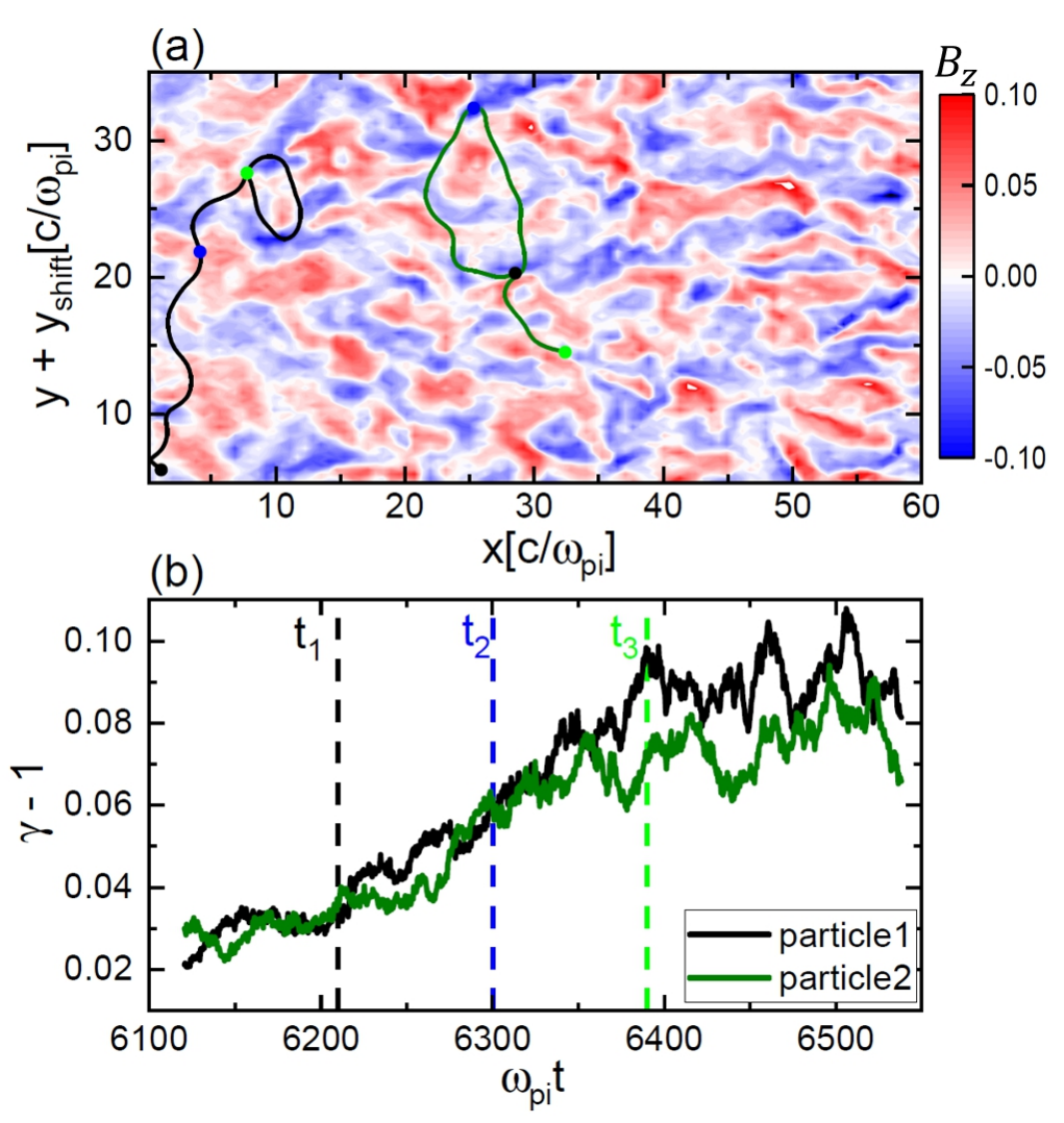}}
\vskip -0.2cm
\caption{(a) Trajectories of two selected electrons overlaid on the 2D distribution of $B_z$ (normalized by $\sqrt{8\pi \mathcal{E}_{\rm kin}}$) at $t_3$. 
The dots indicate the positions at the three specific epochs: $ t_1 \sim 6210 \omega_{\rm pi}^{-1}$ (black), $t_2 \sim 6300 \omega_{\rm pi}^{-1}$ (blue), and $t_3 \sim 6390 \omega_{\rm pi}^{-1}$ (green). 
(b) Time evolution of the electron Lorentz factor.  Note that the shock propagation distance, $u_{\rm sh}^{\rm sim}(t_3 - t_1) \sim 0.8 c/\omega_{\rm pi}$, is much smaller than the traveling distances of the electrons. \label{f3}}
\end{figure}

\begin{figure}[t]
\vskip 0.2 cm
\hskip 0.0 cm
\centerline{\includegraphics[width=0.5\textwidth]{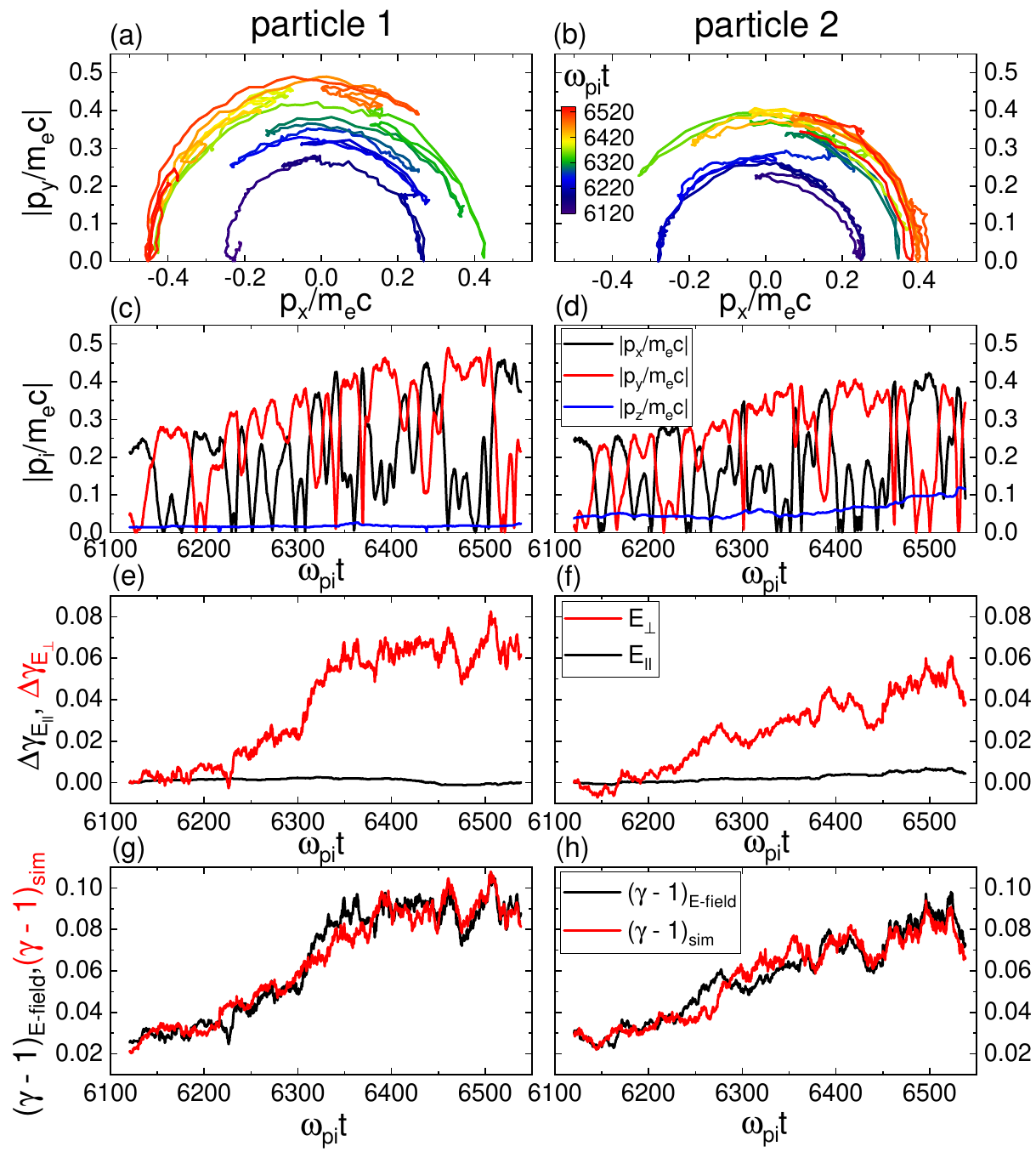}}
\vskip 0.0 cm
\caption{(a)-(b) Trajectories of the two selected electrons shown in Figure \ref{f3} in the $p_x$-$p_y$ space. 
(c)-(d) Three momentum components, $p_j$, where the subscript stands for $j = x$, $y$, or $z$.  
(e)-(f) Energy gain via the two electric field components either parallel or perpendicular to the local magnetic field, $\Delta \gamma_{\rm E_{\parallel}}$ (black) and $\Delta \gamma_{\rm E_{\perp}}$ (red).
(g)-(h) The red lines show $\gamma_{\rm sim} - 1$ in the simulation frame, while the black lines show $\gamma_{\rm E-field}-1$. 
The left and right columns are for the particle 1 and 2, respectively.  \label{f4}}
\end{figure}

We first attempt to understand the main preacceleration process in our simulations.
In comparison with the aforementioned studies of high-$M_A$ shocks with $\beta\lesssim 1$,
we find the following differences.
(1) The 1D profile of $\langle n_i \rangle_y$ averaged along the $y$-direction shown in Figure \ref{f1}(e) 
exhibits a broad and smooth transition without a steep ramp or overshoot/undershoot oscillations. 
(2) Both SDA and SSA are expected to be ineffective in our simulations because of small $B_0$.
Moreover, the gradient drift due to the compression of transverse magnetic fields at the shock transition should be suppressed, since the fluctuating fields dominate over any coherent fields.
(3) The Weibel-excited turbulent magnetic fields dominate over the initial magnetic field (i.e., $\delta B_z \gg B_0$), hence the electric field fluctuations are randomized in the shock transition.
As a result, we expect the stochastic Fermi II acceleration via pitch-angle diffusion by interactions with the Weibel-induced turbulence is the dominant preacceleration process in our simulations.
(4) Magnetic reconnection is unlikely to make a significant contribution to preacceleration in the high-$\beta$ regime, because the magnetic energy density is much smaller than the bulk kinetic energy density.
 
Figure \ref{f3} shows the trajectories of two selected electrons and the time evolution of their Lorentz factor, $\gamma$, during the preacceleration stage. 
The electrons indeed interact multiple times with turbulent magnetic fluctuations in the shock transition, while they gain energy steadily and stochastically during the period of $\sim t_1 - t_3$. 
This is consistent with the 2D PIC simulations of \citet{bohdan2017} (see also their Figure 13(e)).

\begin{figure}[t]
\vskip 0.0 cm
\hskip -0.2 cm
\centerline{\includegraphics[width=0.45\textwidth]{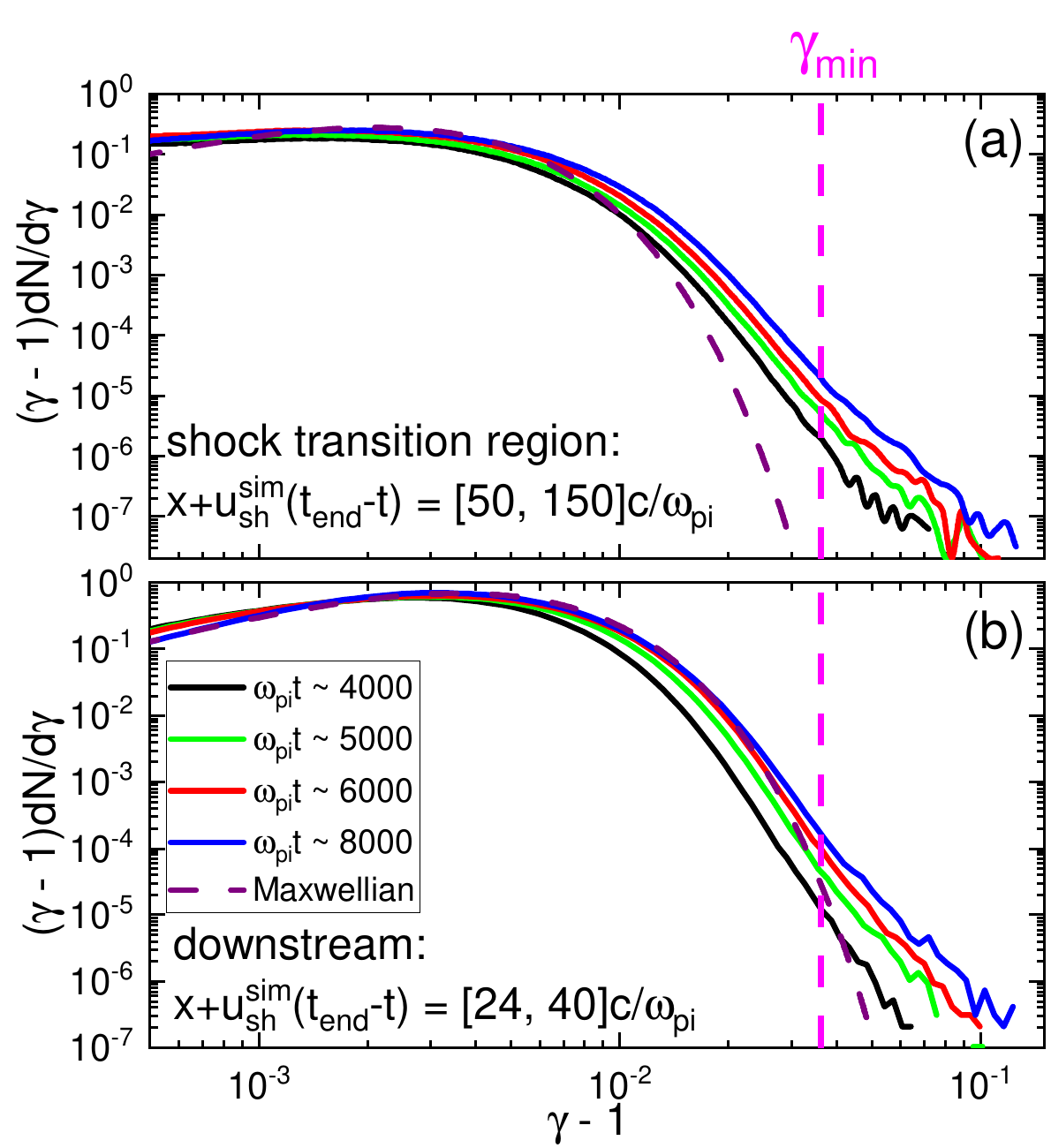}}
\vskip -0.2cm
\caption{Electron energy spectra measured in the shock transition region (top) and the downstream region (bottom) at the four different time epochs. The purple dashed line shows the Maxwellian distribution for the temperature estimated in the corresponding region. The vertical magenta line corresponds to $p_{\rm min}\approx 0.26 m_ec$ ($\gamma_{\rm min}-1=0.036$). Note that $\gamma_{\rm inj}\approx 4.6$ lies outside the upper boundary of the plot. The shock transition region is the same as the one shown in Figure \ref{f2}.\label{f5}}
\end{figure}

To further examine how these two electrons gain energy,
we display their trajectories in the $p_x$-$p_y$ plane in Figure \ref{f4}(a)-(b).
Multiple arcs of electron trajectories in the momentum space imply that they gain energy via multiple interactions with turbulent magnetic waves, while undergoing pitch-angle diffusion.  
This is the characteristic feature of stochastic Fermi II acceleration process.  
In Figure \ref{f4}(c)-(d), we also show the time evolution of $p_x$ (black), $p_y$ (red), and $p_z$ (blue) of the two electrons,
indicating they gain momentum stochastically in the simulation plane.

In Figure \ref{f4}(e)-(f), we show the energy gains due to the work done by the electric field components either parallel ($E_{\parallel}$) or perpendicular ($E_{\perp}$) to the local magnetic field:
 $\Delta \gamma_{\rm E_{\parallel}} \equiv - \int e v_{\parallel} E_{\parallel}dt$
and $\Delta \gamma_{\rm E_{\perp}} \equiv - \int e v_{\perp} E_{\perp}dt$.
The figure demonstrates that the electrons are energized mainly through the interactions with $E_{\perp}$ (red), corroborating that the stochastic Fermi II process operates here. 
In Figure \ref{f4}(g)-(h), the time evolution of the kinetic energy, $(\gamma-1)_{\rm E-field}$ (black), estimated with the electric field is compared with that of $(\gamma-1)_{\rm sim}$ in the actual simulation data.
This exercise confirms that in the simulations electrons gain energy through the work done by the turbulent electromagnetic fluctuations generated by the Weibel instability. 

Figure \ref{f5} illustrates that the electron energy spectra measured in the shock transition region and the downstream region exhibit suprathermal tails for $p\gtrsim p_{\rm min}$,  
where $p_{\rm min} \approx p_{\rm inj}(m_e/m_i)^{1/2}$ is the smallest momentum above which the energy spectrum deviates from the Maxwellian. 
The number fraction of such suprathermal electrons in the shock downstream is roughly $\sim 10^{-4}$ at $t_{\rm end}$.
For the further energization of suprathermal electrons up to $p_{\rm inj}$, multi-scale waves with wavelengths comparable to the electron gyroradius for $p_{\rm inj}$, $\lambda \sim r_{\rm ge}(p_{\rm inj})$, must be present in the shock transition. 
In the fiducial model, $r_{\rm ge}(p_{\rm inj}) \sim 12 c/\omega_{\rm pi}$,
whereas the Weibel-induced filaments extend to $\lambda_{\rm max} \sim 10 c/\omega_{\rm pi}$.
Indeed, the maximum electron energy in the energy spectra increases with time in Figure \ref{f5}. 
Therefore, we expect that electrons could be preaccelerated and injected into DSA via stochastic Fermi II acceleration on sufficiently long timescales, thanks to the ion-scale fluctuations induced by the ion-beam Weibel instability at high-$M_A$ shocks considered here. 

\begin{figure}[t]
\vskip 0.0 cm
\hskip -0.2 cm
\centerline{\includegraphics[width=0.45\textwidth]{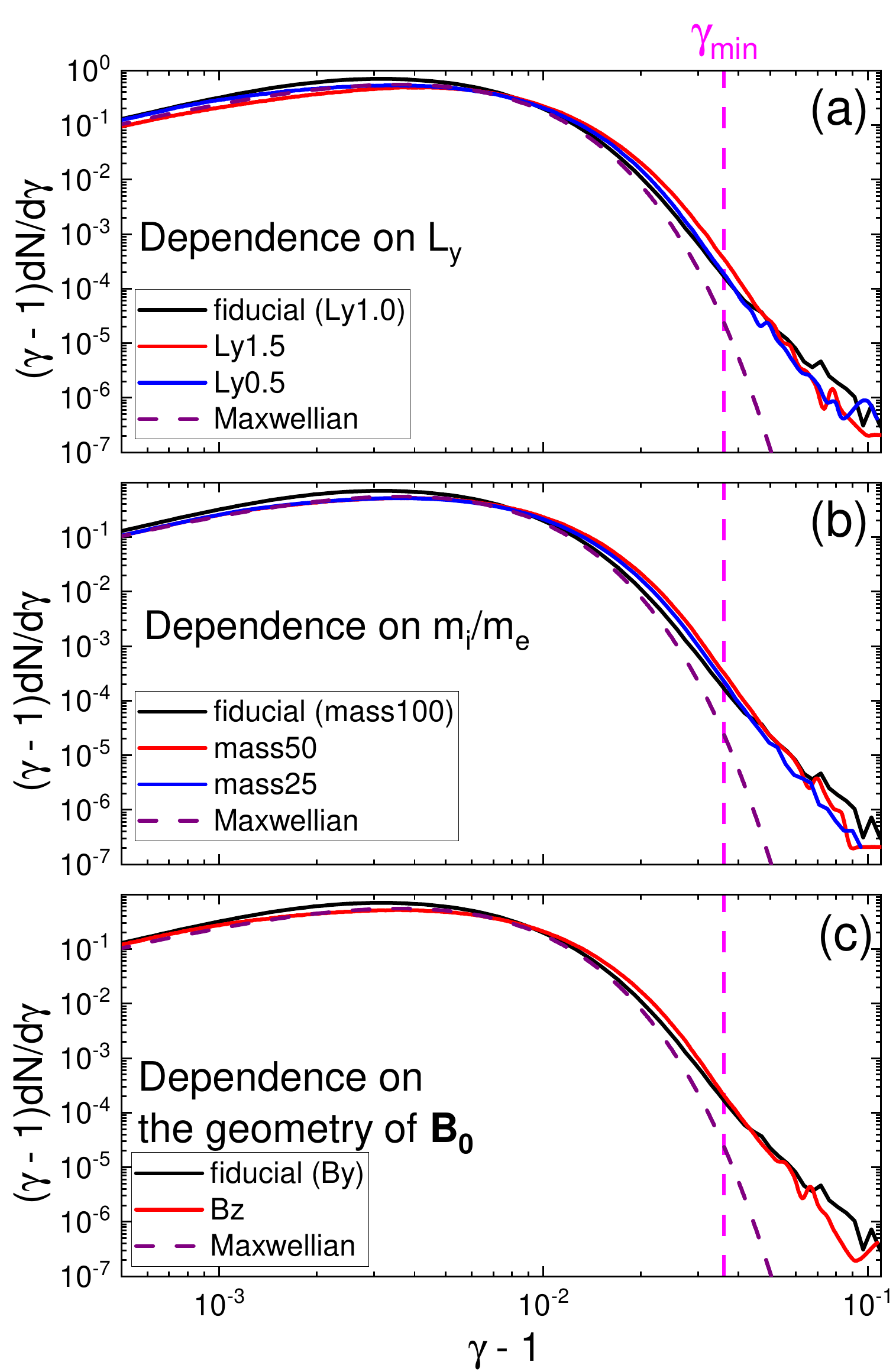}}
\vskip -0.2cm
\caption{Electron energy spectra measured in the shock downstream region at $\omega_{\rm pi}t \sim 8.0 \times 10^3$:
(a) with different transverse box size, $L_y$ (black), $1.5L_y$ (red), and $0.5L_y$ (blue), (b) with different mass ratio, $m_i/m_e=100$ (black), 50 (red), and 25 (blue), and (c) with different geometries of $\bf{B_0}$, in-plane (black) and out-of-plane (red). The vertical magenta lines correspond to $\gamma_{\rm min}$.} \label{f6}
\end{figure}

\begin{figure}[t]
\vskip 0.0 cm
\hskip -0.2 cm
\centerline{\includegraphics[width=0.45\textwidth]{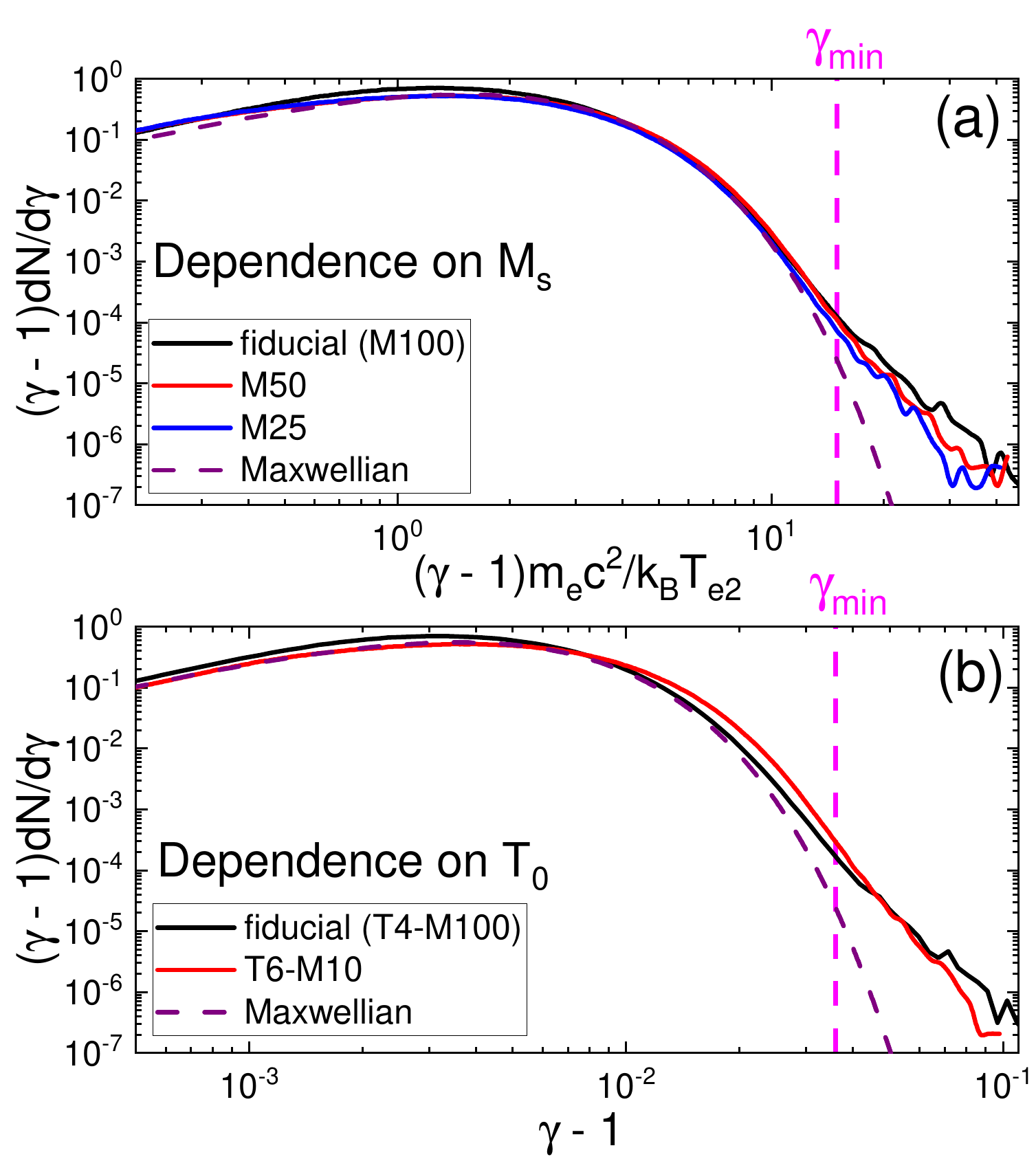}}
\vskip -0.2cm
\caption{Electron energy spectra measured in the shock downstream region at $\omega_{\rm pi}t \sim 8.0 \times 10^3$:
(a) with different Mach number, $M_s=100$ (black), 50 (red), and 25 (blue), and (b) with different preshock temperature, $T_0=10^4$~K ( $M_s=100$, black) and $T_0=10^6$~K ($M_s=10$, red). 
In panel (a), the $(\gamma-1)$ factor is re-scaled with the postshock electron thermal energy (i.e., $k_BT_{e2}/m_ec^2$). The vertical magenta lines correspond to $\gamma_{\rm min}$.\label{f7}}
\end{figure}

\subsection{Dependence on Model Parameters}
\label{s2.4}

Previous studies explored the dependence of the Buneman and Weibel instabilities on the mass ratio and the configuration of the background magnetic field \citep[e.g.,][]{bohdan2017, bohdan2019a, bohdan2019b}.
They found that 2D PIC simulations with an out-of-plane magnetic field
better reproduce the overall picture of the Buneman instability and electron SSA, compared to those with an in-plane magnetic field.
However, simulations can follow reasonably well the Weibel instability and stochastic Fermi acceleration with the 2D in-plane configuration.
The initial energization occurs through SSA, although it becomes less significant in simulations with higher $m_i/m_e$.
Since the Weibel instability is the dominant microinstability and the background field $B_0$ is much weaker than the Weibel-induced magnetic turbulence in the simulations considered here,
we expect the overall results of our study would not strongly depend on the mass ratio nor on the initial magnetic field configuration.

Yet, in this section, we examine how electron preacceleration depends on simulation parameters, such as $L_y$, $m_i/m_e$, and the geometry of $\mathbf{B_0}$, and also on the shock properties, such as $M_s$ and $T_0$ (see Table \ref{t1}). 
The electron energy spectra measured in the shock downstream in simulations with different parameters are shown in Figures \ref{f6} and \ref{f7}. 
Figure \ref{f6}(a) demonstrates that the acceleration process operates efficiently regardless of $L_y$,
as long as the transverse size is sufficiently large to cover the mean separation of Weibel-induced filaments. 
The length scales of the shock transition and ion-Weibel filaments are governed by the ion skip depth, and hence the acceleration does not depends strongly on the mass ratio $m_i/m_e$ (Figure \ref{f6}(b)). 
As expected, the downstream energy spectrum is almost independent of the geometry of $\mathbf{B_0}$ (Figure \ref{f6}(c)).

Moreover, Figure \ref{f7} demonstrates that electron preacceleration is similarly effective for accretion shocks with broad ranges of shock parameters, which are induced by the accretion from voids ($T \sim 10^4$ K, $M_s = 10 - 10^2$) and filaments ($T \sim 10^6$ K, $M_s \lesssim 10$). 
In panel (a), the energy factor, $(\gamma-1)$, is re-scaled with the ratio of $k_B T_{e2}/m_e c^2$, since the postshock electron temperature increases with $M_s$ as $T_{e2}\propto M_s^2$.
We find that the electron to ion temperature ratio is typically $T_{e2}/T_{i2}\sim 0.3$ in the postshock region in our PIC simulations. However, we expect that the temperature equilibrium would be eventually established in the far downstream, so we assume $T_{e2}\approx T_{i2}$ for the thermal leakage injection model to be described in Section \ref{s3}. 

In our 2D PIC simulations, the wave vectors are confined in the $x$-$y$ domain, so the energy gain via $E_z$ is negligible as shown in Figure \ref{f4}(e)-(f). 
In 3D simulations, however, electrons could be energized through the work done by all three components of the electric field owing to the presence of 3D structures of the Weibel filaments.
Despite such limitations of 2D simulations, the Weibel instability is triggered as long as $T_{i\perp} > T_{i\parallel}$. 
Hence, the acceleration process mediated by the Weibel-induced waves should operate regardless of the simulation dimensionality.
For instance, \citet{matsumoto2017} showed that 2D simulations with an in-plane initial magnetic field provide the realistic 3D picture of the Weibel instability and the second-order Fermi-like acceleration process.

In summary, the 2D PIC simulations of comparison models presented in this section imply that electrons could be preaccelerated and injected to DSA at high-$M_A$ shocks even in the low magnetization regime of accretion shocks.
Therefore, in the discussion below, we presume that the canonical DSA power-law spectrum forms in the downstream region of strong accretion shocks.

We note that we did not observe any significant pre-energization of suprathermal protons during $t \lesssim 8 \times 10^3 \omega_{\rm pi}^{-1}$ in our PIC simulations. However, we expect that proton preacceleration and injection via stochastic Fermi II acceleration would ensue on sufficiently longer timescales, since the maximum wavelength of the ion-Weibel-induced turbulence is comparable to the gyroradius of protons with $p_{\rm inj}$. 

\section{Analytic DSA Model for CR Spectra in Test-particle Regime}
\label{s3}

Although DSA is relatively well established, the preacceleration of thermal particles up to $p_{\rm inj}$ has not been fully understood.
Before high capacity hybrid and PIC simulations have become feasible, some earlier studies adopted a phenomenological model for ``thermal leakage injection'', in which particles above a certain injection momentum cross the shock front and get injected into the CR population \citep[e.g.][]{malkov1998, gieseler2000, kang2002}.
In the test-particle regime, the proton momentum distribution is described as the combination of the postshock Maxwellian distribution and the DSA power-law, $f_{\rm CRp} \propto p^{-q}$ (where $q = 4M_{\rm s}^2/(M_{\rm s}^2-1)$), extending from $p_{\rm inj}$ to higher momenta \citep{kang2010,ryu2019}.

Based on hybrid simulations with kinetic proton population, a series of papers by \citet{caprioli2014a,caprioli2014b} demonstrated that at strong $\beta \approx 1$ shocks, the postshock proton distribution is composed of three components, Maxwellian, suprathermal bridge, and non-thermal power-law.
In the late stage, the far-downstream proton spectrum approaches to the combination of the Maxwellian and the power-law with the DSA slope. 
Hence, the classical thermal leakage model for the CRp injection has acquired strong confirmation from these hybrid simulations.

As described in the introduction, in the last decade or so, many studies using PIC simulations demonstrated that at high-$\beta$ $Q_{\perp}$ shocks electrons could be preaccelerated owing to the multi-scale waves excited by various microinstabilities, and that suprathermal electrons form a power-law distribution, extending from a certain minimum momentum, $p_{\rm min}$, to higher momenta. 
Considering those earlier studies, \citet{kang2020} suggested that the far-downstream power-law spectrum of CRe could be constructed from a thermal leakage model in the manner analogous to that of CRp.
The electron power-law spectrum, $f_{\rm CRe} \propto p^{-q}$, extends from a Maxwellian, starting from $p_{\min}\sim 3.5 p_{\rm th,e}\ll p_{\rm inj}$.
In fact, \citet{arbutina2021} suggested injection recipes for CRp and CRe at both $Q_{\parallel}$ and $Q_{\perp}$ shocks, which are in line with such expectations, by performing large-scale PIC simulations for high-Mach-number, low-$\beta$, parallel shocks.
Their simulations corroborated that in the far-downstream region both CRp and CRe populations develop power-law spectra with a slope similar to the DSA slope $q$, and that the lowest momenta of the power-law spectra satisfy the condition $p_{\rm inj}/p_{\rm th,p} \approx p_{\rm min}/p_{\rm th,e} \sim 3.7$, in the test-particle regime with a weak shock modification.

Therefore, for the external accretion shocks considered in the previous section as well as for weak internal shocks (see Table \ref{t1}), we adopt a prescription, in which nonthermal protons develop a test-particle power-law spectrum:
\begin{equation}
f_{\rm CRp}(p) \approx f_{\rm inj} \cdot \left(\frac{p}{p_{\rm inj}}\right)^{-q}{\exp}\left(-{p^2 \over p_{\rm max}^2}\right)~~{\rm for}~p\geq p_{\rm inj},
\label{crp}
\end{equation} 
where $p_{\rm inj} = Q\cdot p_{\rm th,p}$ is injection momentum with $Q\sim 3.5-4.0$ \citep{ryu2019}. 
The normalization factor, $f_{\rm inj} = n_2 \pi^{-1.5} p_{\rm th,p}^{-3} {\rm exp}(-Q^{2})$, is specified by the Maxwellian distribution, where $n_2$ is the postshock gas density.
Note that the highest momentum of CRp, $p_{\rm max}\gg m_pc$, is governed by several unspecified elements, such as the shock age, magnetic field fluctuations, and energy losses, which are beyond the scope of this study.
Moreover, in strong accretion shocks, the nonlinear DSA effects could become important, depending on the injection parameter $Q$ and $p_{\rm max}$,
and hence the single power-law spectrum may need to be modified \citep[e.g.][]{malkov2001,caprioli2010,arbutina2021}.
However, the upper end of $f_{\rm CRp}$ does not much affect the estimation of the nonthermal emission due to CRp-p inelastic collisions, which will be presented in the next section.
 
\begin{figure}[t]
\vskip 0.0 cm
\hskip -0.2 cm
\centerline{\includegraphics[width=0.55\textwidth]{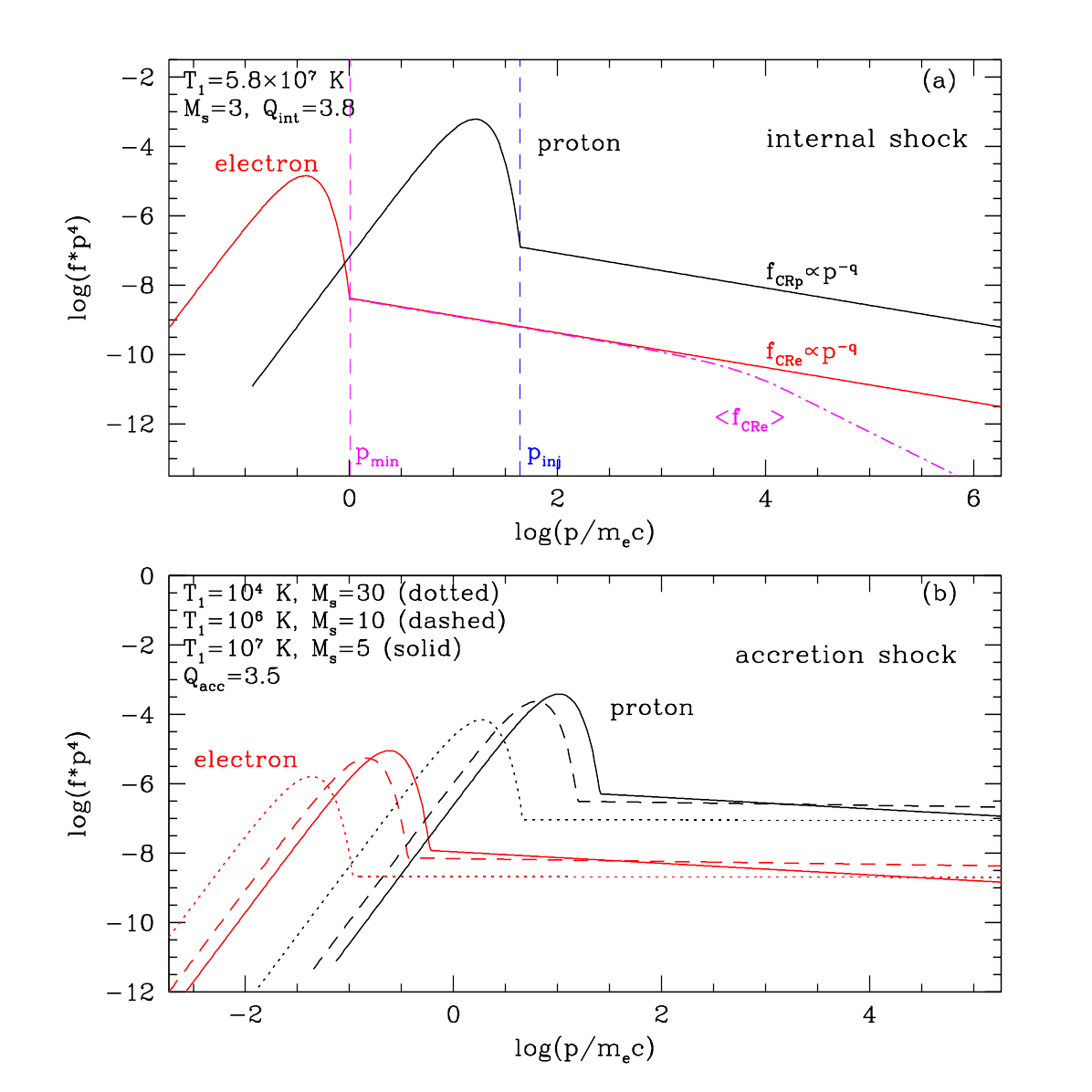}}
\vskip -0.2cm
\caption{Maxwellian distributions along with the DSA power-law spectra, $f_{\rm CRp}$ (black) and $f_{\rm CRe}$ (red), for protons and electrons, respectively, in the postshock region.
The model parameters are specified in each panel.
(a) Internal shock with $T_1=5.8\times 10^7$~K and $M_s=3$.
The blue vertical line corresponds to $p_{\rm inj} = Q_{\rm int}\cdot p_{\rm th,p}$, while the magenta vertical line corresponds to $p_{\rm min} = Q_{\rm int}\cdot p_{\rm th,e}$. Here, $Q_{\rm int}=3.8$.
 The magenta dot-dashed line shows the volume-averaged electron spectrum, $\langle f_{\rm CRe}\rangle$, calculated for $t_{\rm adv}\approx 54$~Myr, $B_2=1\mu {\rm G}$, and $z_r=0$. 
(b) Accretion shocks with $T_1=10^4-10^7$~K, $M_s=5-30$, and $Q_{\rm acc}=3.5$. \label{f8}}
\end{figure}

Similarly, the spectrum for nonthermal electrons may be expressed as
\begin{equation}
f_{\rm CRe}(p) \approx f_{\rm min} \cdot \left({p \over p_{\rm min}}\right)^{-q}{\exp}\left(-{p^2 \over p_{\rm eq}^2}\right) ~~{\rm for}~p\geq p_{\rm min},
\label{cre}
\end{equation} 
where $p_{\rm min} = Q\cdot p_{\rm th,e}$ stands for the smallest electron momentum above which the suprathermal power-law tail develops \citep{kang2020}.
As mentioned in Section \ref{s2.4}, we assume $T_2=T_{i2}=T_{e2}$ in the far downstream region, so both ions and electrons could be described with the Maxwellian distributions of the same postshock temperature.
The normalization factor is prescribed as $f_{\rm min} = n_2 \pi^{-1.5} p_{\rm th,e}^{-3} {\rm exp}(-Q^{2})$.
The cutoff momentum, $p_{\rm eq}$, is fixed by the equilibrium condition that the DSA energy gain per cycle is equal to the synchrotron/IC losses per cycle.  

Panel (a) of Figure \ref{f8} illustrates these power-law DSA spectra for a $M_s=3$, internal shock along with $p_{\rm min}$ and $p_{\rm inj}$,
while panel (b) displays the spectra for three strong shocks with different values of $T_1$ and $M_s$.
By adopting Equation (\ref{cre}), we presume in effect that thermal electrons are energized from $p_{\rm min}$ to $p_{\rm inj}$ via various preacceleration processes such as Fermi-like SDA, SSDA, and Fermi II acceleration, while CR electrons are accelerated via DSA for $p> p_{\rm inj}$. 

Here, for simplicity, the same injection parameter $Q$ is adopted for both CRp and CRe.
Then, the CR injection fraction by number, $\xi_p$ and $\xi_e$, depend only on $Q$ and $q$ as follows \citep{ryu2019}:
\begin{equation}
\xi_{p,e}\equiv \frac{n_{\rm CRp,e}}{n_2} = \frac{4}{\sqrt{\pi}}Q^3 {\rm exp}(-Q^2)\frac{1}{q-3}. 
\label{xipe}
\end{equation}
So the adoption of the same value of $Q$ means that 
the numbers of the nonthermal particles injected to preacceleration and then DSA are the same for both protons and electrons.
For an internal shock with $M_s=3$ ($q=4.5$), for example, 
$\xi_p=\xi_e\approx 4.4\times 10^{-5},~3.1\times 10^{-4},$ and $1.8\times 10^{-3}$ for $Q=3.8, ~3.5$, and $3.2$, respectively.
Furthermore, with the DSA model spectra given in Equations (\ref{crp}) and (\ref{cre}), the ratio of $f_{\rm CRp}$ to $f_{\rm CRe}$ at  $p=p_{\rm inj}$ (i.e., the CRp-to-CRe number ratio) can be calculated as
\begin{equation}
K_{p/e} \equiv \frac{f_{\rm CRp}(p_{\rm inj})}{f_{\rm CRe}(p_{\rm inj})} = \left({p_{\rm th,p} \over p_{\rm th,e}}\right)^{q - 3} = \left({m_{p} \over m_{e}}\right)^{(q - 3)/2}.
\label{Kpe}
\end{equation}
For strong accretion shocks with $q \approx 4$, the ratio is $K_{p/e} \approx 43$ with $m_p/m_e=1836$; it increases as $q$ increases, and hence is larger for internal shocks.

The postshock CRp and CRe energy densities can be estimated as
\begin{equation}
\mathcal{E}_{\rm CRp,e} = 4 \pi c \int_{p_0}^{p_1} (\sqrt{p^2+ (m_{p,e}c)^2}-m_{p,e}c) f_{\rm CRp,e}(p) p^2 dp.
\label{ECR}
\end{equation}
In general, this energy integral depends on the lower and upper bounds, $p_0$ and $p_1$.
For strong shocks with $q=4$, for example, the CR energy density increases with the highest momentum as $\mathcal{E}_{\rm CRp,e}\propto \ln p_1$.
For weaker shocks with $q>4$, on the other hand, $\mathcal{E}_{\rm CRp,e}$ converges to a constant value for $p_1/m_{p,e}c\gg 1$, while
it depends on the minimum energy $p_0$ as well as the slope $q$.
Different values of $p_0$ and $p_1$ have been adopted for the calculation of $\mathcal{E}_{\rm CRp,e}$ in existing literature \citep[e.g.,][]{pinzke2010,vazza2016,ryu2019,wittor2020},
We here adopt $p_0=p_{\rm min}$ and $p_1=p_{\rm eq}$ for CRe and $p_0=p_{\rm inj}$ and $p_1=p_{\rm max}$ for CRp.

The ``acceleration efficiencies'' of CRp and CRe are often defined in terms of the shock kinetic energy flux as follows:
\begin{equation}
\eta_{p,e}\equiv \frac{\mathcal{E}_{\rm CRp,e} u_2}{(1/2) n_1 m_p u_s^3}=\frac{1}{r}\frac{\mathcal{E}_{\rm CRp,e}}{\mathcal{E}_{\rm sh}}
\end{equation}
\citep{ryu2003, ryu2019}.
For an internal shock with $M_s=3$ and $T_1=5.8\times 10^7$~K, our DSA power-law model yields
$\eta_p\approx 2.9 \times 10^{-3},~1.8 \times 10^{-2},$ and $8.9\times 10^{-2} $, and
$\eta_e\approx 1.1 \times 10^{-5},~6.7 \times 10^{-5},$ and $3.3\times 10^{-4} $,
for $Q=3.8, ~3.5,$ and $3.2$, respectively. 
So with $Q\lesssim 3.2$, $\eta_p\gtrsim 0.1$, and the test-particle assumption may need to be revised.
For an accretion shock with $M_s=10^2$, $T_1=10^4$~K, $p_{\rm max}=10^8 m_pc$, and $Q=3.8$, $\eta_p\approx 0.16$ and $\eta_e\approx 2.9 \times 10^{-3}$.
For a shock with $M_s=10$, $T_1=10^6$~K, $p_{\rm max}=10^8 m_pc$, and $Q=3.8$, $\eta_p\approx 0.11$ and $\eta_e\approx 1.7 \times 10^{-3}$.
Therefore, unlike the number fractions, $\xi_p$ and $\xi_e$, the CR acceleration efficiencies, $\eta_p$ and $\eta_e$, depend on $T_1$, $M_s$, and $Q$, as well as $p_0$ and $p_1$, in the DSA model adopted here.
The amplitude, $f_{\rm CRp,e}\propto \exp(-Q^2)Q^q$, and so $\mathcal{E}_{\rm CRp,e}$ and $\eta_{p,e}$ follow approximately the same scaling.
We note that, if the CR pressure becomes significant and hence the nonlinear effects of DSA become important with $\eta_p \gtrsim 0.1$, the postshock temperature decreases accordingly and the ensuing CR spectrum deviates from the test-particle power law \citep{malkov2001}.
In that regime, a kind of self-adjustment such as the scheme employed by \citet{ryu2019} might be relevant.

As CRe accelerated at shocks are advected downstream in the postshock region, they lose energy due to synchrotron/IC losses with the cooling time $t_{\rm cool}(\gamma) \approx 10^8 {\rm yr} (B_{\rm e,2}/5\mu {\rm G})^{-2}(\gamma/10^4)^{-1}$ and the cooling length $l_{\rm cool} \approx 25 {\rm kpc} (B_{\rm e,2}/5\mu {\rm G})^{-2}(\gamma/10^4)^{-1} (u_2/250 {\rm km~s^{-1}})^{-1}$, where again $\gamma$ is the Lorentz factor of CRe. 
Here, the ``effective" magnetic field strength, $B_{\rm e} = \sqrt{B^2 + B_{\rm rad}^2}$ with $B_{\rm rad} = 3.24 (1+z_r) \mu$G, accounts for the loss due to synchrotron as well as IC scattering off CMB, where $z_r$ is the cosmological redshift. 
As a result, the volume-integrated downstream momentum spectrum of CRe steepens by one power of $p$ as $\propto p^{-(q+1)}$ for $p \gtrsim p_{\rm br}$, where the ``break" momentum,
\begin{equation}
p_{\rm br} = 10^4 m_e c \left(t_{\rm adv} \over 10^8 {\rm yr}\right)^{-1} \left(B_{\rm e,2} \over 5\mu {\rm G}\right)^{-2},
\label{pbr}
\end{equation}
is estimated from the condition that the advection timescale, $t_{\rm adv}$, is equal to $t_{\rm cool}$. 

In the next section, we use the volume-averaged CRe spectrum to evaluate synchrotron and IC emission in numerical shock zones identified in structure formation simulations.
To obtain the averaged spectrum in the postshock volume with $l_{\rm adv}=t_{\rm adv}u_2$, we perform the numerical integration of electron cooling along the 1D plane-parallel slab under the assumption of continuous injection \citep{carilli1991} and a constant $B_{\rm e,2}$.
Note that the postshock cooling depends only on $B_{\rm e,2}$, and so this cooling integration becomes self-similar, if we adopt the similarity variable, $\chi \equiv p/p_{\rm br}$.
The volume-averaged spectrum of CRe within each shock zone could be expressed as
\begin{equation}
\langle f_{\rm CRe}\rangle \approx f_{\rm min} \cdot \left({p \over p_{\rm min}}\right)^{-q}{\rm exp}\left(-{p^2 \over p_{\rm eq}^2}\right) \left(1 + \left(\chi \over \chi_0 \right)^a \right)^{-1/a},
\label{fcreva}
\end{equation}
where the fitting parameters are found to be $\chi_0 \approx 0.21$ and $a \approx 1.6$.
The fitting form for the cooling integral, $\left(1 + (\chi / \chi_0)^a \right)^{-1/a}$, could be simplified as $(\chi / \chi_0)^{-1}$ for $\chi \gg 1$, as desired.
Thus, the volume-averaged electron spectrum, $\langle f_{\rm CRe}\rangle$, is determined by $n_2$, $T_2$, $M_s$, $u_2$, $B_2$ of each shock zone as well as its size $\Delta l$ and the redshift $z_r$.
In Figure \ref{f8}, the magenta dot-dashed line draws $\langle f_{\rm CRe}\rangle$ integrated over $t_{\rm adv} = 54$~Myr in the postshock flow with $B_2=1\mu{\rm G}$ and $z_r=0$.

\section{Nonthermal Radiation from the Outer Regions of Galaxy Clusters}
\label{s4}

\subsection{Numerical Shocks in Cosmological Simulations}
\label{s4.1}

\begin{figure*}[t]
\vskip 0.0 cm
\hskip -0.2 cm
\centerline{\includegraphics[width=0.95\textwidth]{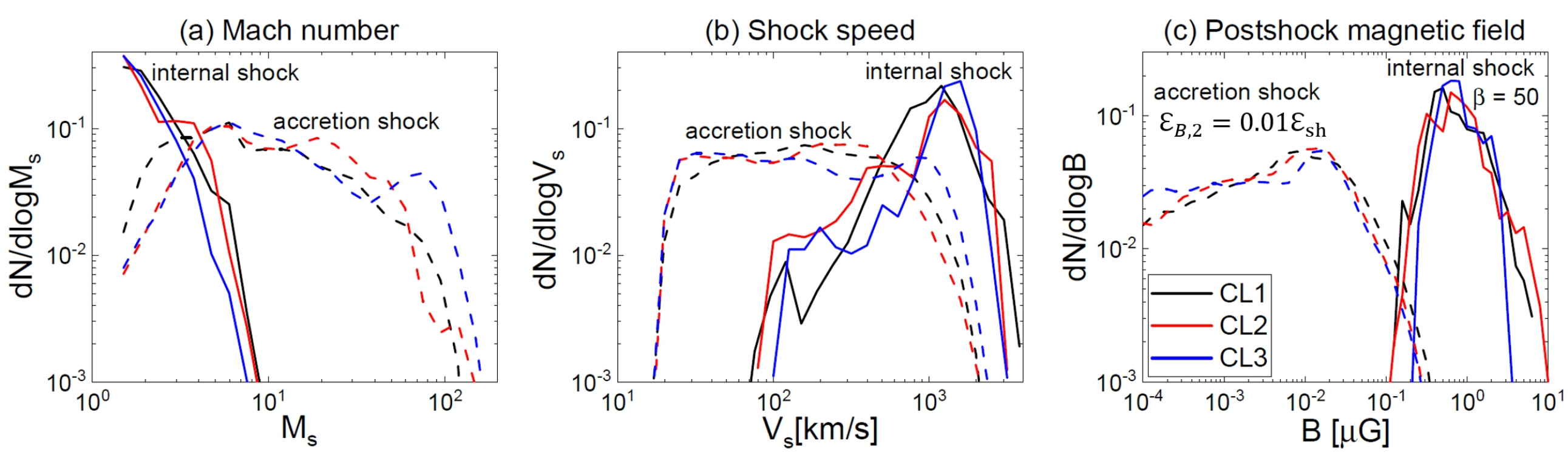}}
\vskip -0.2cm
\caption{Probability distribution functions (PDFs) of the properties of shock zones associated with the three sample clusters, CL1 (black), CL2 (red), and CL3 (blue): (a) the sonic Mach number $M_s$, (b) the shock speed $V_s=M_s\cdot c_{s1}$, and (c) the postshock magnetic field strength $B_2$.  The solid and dashed lines corresponds to the shock zones classified as internal and accretion shocks, respectively. 
For internal shocks the magnetic field strength is assumed to satisfy the condition of $\beta=50$ in the postshock flow.
For accretion shocks, on the other hand, the magnetic field is assumed to be amplified to the level of $\mathcal{E}_{B,2} \approx 0.01 \mathcal{E}_{\rm sh}$.
\label{f9}}
\end{figure*}

We performed a set of cosmological hydrodynamic simulations to obtain synthetic cluster samples using a cosmological structure formation code \citep{ryu1993}, as described in \citet{ha2018a}. The parameters from the WMAP7 data \citep{komatsu2011} were employed: baryon density $\Omega_{\rm BM} = 0.046$, dark matter density $\Omega_{\rm DM} = 0.234$, cosmological constant $\Omega_{\Lambda} = 0.72$, Hubble parameter $h \equiv H_0/(100 {\rm km s^{-1} Mpc^{-1}}) = 0.7$, rms density fluctuation $\sigma_{8} = 0.82$, and primordial spectral index $n = 0.96$. The simulation box has the comoving size of $100 h^{-1}$ Mpc with $2048^3$ grid zones and periodic boundaries, so the spatial resolution is $\Delta l = 48.8 h^{-1}$ kpc. Because nongravitational effects such as radiative and feedback processes are expected to be not so important in the outskirts of galaxy clusters, they were not included in the simulations. 

In the simulation box, the centers of clusters are identified as the local peaks of X-ray emissivity, and the X-ray emission-weighted temperature inside $r_{200}$, $T_X$, is estimated  for each cluster. 
Here, $r_{200}$ is the virial radius defined by the gas overdensity of $\rho_{\rm gas}/\left<\rho_{\rm gas}\right> = 200$. 
For this paper, three sample clusters are selected: CL1 with $kT_{X,1} \approx 4.8$ keV and $r_{200,1} \approx 1.25 h^{-1}$ Mpc,  CL2 with $kT_{X,2} \approx 5.1$ keV and $r_{200,1} \approx 1.4 h^{-1}$ Mpc, and CL3 with $kT_{X,3} \approx 5.1$ keV and $r_{200,1} \approx 1.5 h^{-1}$ Mpc.

To find ``shock zones'' in the simulated clusters, we employ the shock identification algorithm described in our previous works \citep[e.g.,][]{ryu2003, ha2018a}. 
In the analysis below, we differentiate the two regions of clusters: the {\it inner region} of $r\le r_{200}$ and the {\it outer region} of  $r_{200}<r <4 r_{200}$. 
The shock zones in the inner region are identified as {\it internal shocks}, while those in the outer region are classified as {\it accretion shocks}.

\subsection{Prescriptions for Magnetic Field Amplification}
\label{s4.2}

While our cosmological simulations are basically hydrodynamic with only ``passive'' magnetic fields, the magnetic field is required for the estimation of nonthermal emission from galaxy clusters.
In particular, the postshock field strength, $B_2$, is necessary to calculate $\langle f_{\rm CRe}\rangle$ and synchrotron radiation.
So we adopt physically-motivated models to specify $B_2$ at each shock zone as follows.
For the shock zones classified as internal shocks in the ICM, $B_2$, is specified by the condition of $\beta=50$ in the postshock flow.
For the shock zones classified as accretion shocks, on the other hand, 
$B_2$ is specified by the condition of $\mathcal{E}_{B,2}= \epsilon_B \mathcal{E}_{\rm sh}$,
where $\mathcal{E}_{B,2}=B_2^2/8\pi$.
As mentioned in Section \ref{s2.2}, the Weibel-amplified field alone may not explain the postshock field on MHD scales.
Nevertheless, we assume that $\epsilon_{B}\approx 0.01-0.1$ could result from various instabilities, turbulent dynamo processes, etc.
The fiducial model adopts $\epsilon_{B}\approx 0.01$, while a comparison model with $\epsilon_{B}\approx 0.1$ is considered to explore the parameter dependence.

Figure \ref{f9} shows the probability distribution functions (PDFs) of the sonic Mach number, $M_s$, the shock speed, $V_s=M_s\cdot c_{s1}$, and $B_2$ of shock zones associated with the three sample clusters.
The $\beta = 50$ recipe applied to internal shocks (solid lines) results in $B_2\approx 0.1-10~\mu$G in the inner region of the clusters.
For accretion shocks (dashed lines), with $\mathcal{E}_{B,2}= 0.01 \mathcal{E}_{\rm sh}$, 
$B_2$ has a broader distribution in the range of $ \sim 10^{-4} - 10^{-1}~\mu$G, with the peak at $\sim10^{-2}~\mu$G. 
These estimates are in a reasonable agreement with the predicted magnetic field strength around cluster regions \citep[e.g.,][]{ryu2008}.  

\begin{figure*}[t]
\vskip 0.0 cm
\hskip -0.2 cm
\centerline{\includegraphics[width=0.9\textwidth]{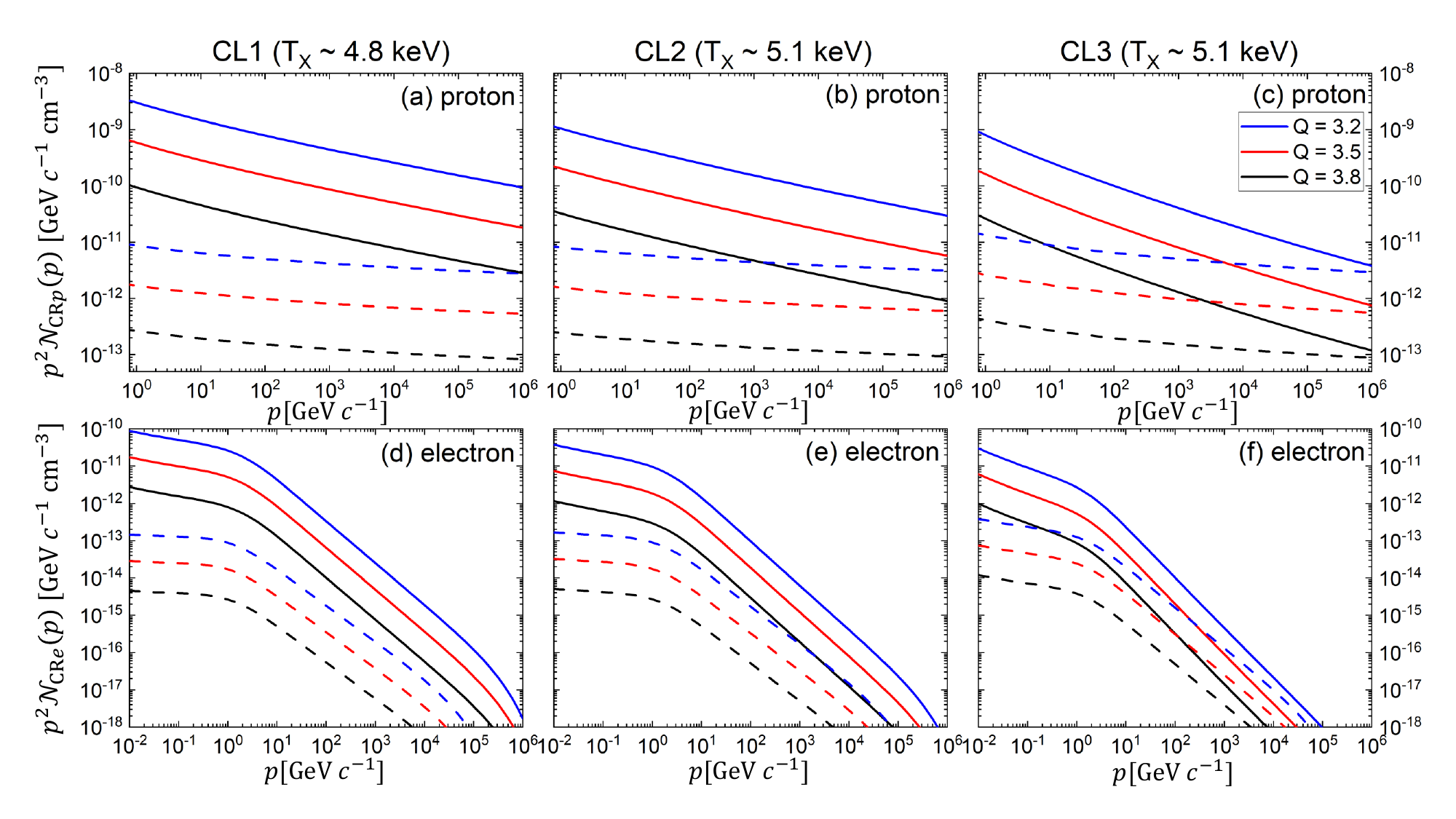}}
\vskip -0.2cm
\caption{Integrated CRp (top) and CRe(bottom) number spectra, $p^2 \mathcal{N}_{\rm int,CRp,e}$ for all internal shocks (solid lines) and $p^2 \mathcal{N}_{\rm acc,CRp,e}$ for all accretion shocks (dashed lines) associated with three sample clusters. 
They are normalized by the total postshock volume of all internal or accretion shock zones associated with each sample cluster, $\sum s_{\rm sh} u_2 \tau_{\rm sh}$ (see Equations (\ref{NCRpint}) and (\ref{NCRpacc})).
The estimates with different values of the injection parameter, $Q=3.2,~3.5$, and $3.8$, are shown in different colors. \label{f10}}
\end{figure*}

\subsection{CR production at Shocks in Simulated Clusters}
\label{s4.3}

In this section, we assess CR production at shock zones associated with the three sample clusters.
As mentioned earlier, at weak internal shocks, CR injection and acceleration depend on the magnetic field obliquity: i.e., protons are accelerated more efficiently at $Q_{\parallel}$ shocks, while electrons are accelerated preferentially at $Q_{\perp}$ shocks (see Table 1).
For accretion shocks, however, background magnetic fields are very weak, and so the magnetic field obliquity is likely to be irrelevant.
Here, we assume both protons and electrons are accelerated at all shock zones regardless of the shock obliquity,
since we do not accurately obtain the magnetic field orientation from our cosmological hydrodynamic simulations.
With that caveat, we ascribe the DSA power-law energy spectra of CRs to each shock zone in the simulation data at $z_r=0$.

To examine how CR production depends on the overall properties of clusters, we calculate the {\it integrated} CRp number spectra,
$\mathcal{N}_{\rm int,CRp}$ and $\mathcal{N}_{\rm acc,CRp}$, at internal and accretion shocks, respectively, by summing up the contributions from all shock zones associated with each sample cluster: 
\begin{equation}
\mathcal{N}_{\rm int, CRp}(p) \approx \frac{\sum \limits_{r_{\rm sh} \leq r_{200}} 4\pi p^2 f_{\rm CRp}  s_{\rm sh}u_2 \tau_{\rm sh}}{\sum \limits_{r_{\rm sh} \leq r_{200}} s_{\rm sh} u_2 \tau_{\rm sh}},
\label{NCRpint}
\end{equation}
and
\begin{equation}
\mathcal{N}_{\rm acc, CRp}(p) \approx \frac{\sum \limits_{r_{200} < r_{\rm sh} \leq 4 r_{200}} 4\pi p^2  f_{\rm CRp} s_{\rm sh}u_2 \tau_{\rm sh}}{\sum \limits_{r_{200} < r_{\rm sh} \leq 4 r_{200}} s_{\rm sh} u_2 \tau_{\rm sh}},
\label{NCRpacc}
\end{equation}
where $s_{\rm sh} = 1.19 (\Delta l)^2$ is the mean surface area of shock zones with random orientations
and $\tau_{\rm sh}$ is the lifetime of the shock.
These integrated spectra are normalized by the total postshock volume of all internal or accretion shock zones associated with each cluster, $\sum s_{\rm sh} u_2 \tau_{\rm sh}$.
To obtain an approximate estimation, 
instead of following the dynamical evolution of individual shock zones,
we assume a uniform value of $\tau_{\rm sh}$ for all shock zones.
Then, $\mathcal{N}_{\rm int,CRp}$ and $\mathcal{N}_{\rm acc,CRp}$ in Equations (\ref{NCRpint}) and (\ref{NCRpacc}) become independent of the value of $\tau_{\rm sh}$. 

The {\it integrated} CRe number spectra, $\mathcal{N}_{\rm int,CRe}$ and $\mathcal{N}_{\rm acc,CRe}$, can be estimated in the same manner by replacing $f_{\rm CRp}$
with $\langle f_{\rm CRe}\rangle$ in Equations (\ref{NCRpint}) and (\ref{NCRpacc}). 
Figure \ref{f10} shows the integrated CRp and CRe spectra for the three sample clusters. 
We include a wide range of $Q \approx 3.2 - 3.8$.
As noted before, $f_{\rm CRp,e}\propto \exp(-Q^2)Q^q$, and so $\mathcal{N}_{\rm CRp,e}$ should follow the same dependence on $Q$.
Several points are noted. 
(1) Although the three clusters have the similar X-ray weighted temperature, 
$\mathcal{N}_{\rm int, CRp,e}$ for internal shocks (solid lines) vary a lot, while $\mathcal{N}_{\rm acc, CRp,e}$ for accretion shocks (dashed lines) are rather similar.
These clusters might have undergone different dynamical evolutions induced by numerous merger events, resulting in diverse histories of shock formation and CR acceleration, especially inside the virial radius. 
On the other hand, gas accretion from voids and filaments does not strongly depend on the internal structures of each cluster,
so CR acceleration at accretion shocks is expected to depend only weakly on the detailed dynamical history of the cluster. 
(2) In the inner region of clusters, internal shocks with $M_s \sim 3 - 4$ dominate in the CR acceleration \citep[e.g.,][]{hong2014, ha2018a}, so the integrated proton spectrum due to internal shocks has $\mathcal{N}_{\rm int,CRp} \propto p^{2-q}$ with $q \sim 4.25 - 4.5$. By contrast, strong accretion shocks produce a flatter integrated CRp spectrum, $\mathcal{N}_{\rm acc,CRp}$ with $q \approx 4 $.
(3) The amplitude of $\mathcal{N}_{\rm acc,CRp,e}$ is much smaller than that of $\mathcal{N}_{\rm int,CRp,e}$, because accretion shocks form in the intergalactic medium with much lower gas density.
(4) The break momentum, $p_{\rm br}$, of the integrated electron spectrum of accretion shocks tends to be higher than that of internal shocks, because on average $B_2$ is much lower at accretion shocks. 

\subsection{Nonthermal Radiation}
\label{s4.4}

Next, we present the estimation of nonthermal emission due to CRe and CRp.
Numerical shock zones are specified by several parameters, such as $n_2$, $T_2$, $M_s$, $u_2$, $B_2$, $f_{\rm CRp}$, and $\langle f_{\rm CRe}\rangle$.
The redshift $z_r$ determines the CMB radiation field and hence the IC cooling rate.
We adopt $z_r = 0$ to focus on nearby clusters such as the Coma cluster. 

The two main parameters that control nonthermal emission in our model calculations are the injection parameter, $Q$, and
the magnetic field parameter, $\epsilon_B$, for accretion shocks.
Obviously, the nonthermal emission scale with the amplitude of CRe and CRp spectra, which in turn scale as $\propto \exp(-Q^2)Q^q$ in our analytic DSA models. 
Since the preacceleration processes for DSA has yet to be fully understood and could be different for weak internal shocks and strong accretion shocks,
below we present the estimates with different values of the injection parameter: i.e., $Q_{\rm int}=3.5$ and $3.8$ for internal shocks and $Q_{\rm acc}=3.2-3.8$ for accretion shocks.
We consider a narrow range of $Q_{\rm int}$ for internal shocks, since $Q_{\rm int}=3.2$ would result in the synchrotron emission from galaxy clusters that are not consistent with observations (see Section \ref{s4.5}).

The synchrotron emission at accretion shocks depends directly on $\epsilon_B$, since the emissivity scales as $\propto B_2^{(q-1)/2}$. 
However, the value of $\epsilon_B$ affects only marginally the CRe spectrum, $\langle f_{\rm CRe}\rangle$,
because with $B_2 \sim 0.01-0.1 \mu {\rm G} \ll B_{\rm rad}$ at accretion shock, CRe cool mostly via IC. 
Hence, the electron IC emission depends rather weakly on $B_2$.
Contrastingly, the $\pi^0$-decay emission is independent of the magnetic field prescription.

We calculate synchrotron, IC, and $\pi^0$-decay emission with the ``one-zone approximation'', using CRe and CRp inside each shock zone and sum them over the volumes of internal and accretion shocks.
Below, for emulation of observations, we use the conventional terminology and units for various physical variables.
For example, the volume emissivity, $j(\nu)$ (${\rm erg~cm^{-3}~s^{-1}~Hz^{-1}~str^{-1}}$), is calculated from $\langle f_{\rm CRe}\rangle$ and $f_{\rm CRp}$ at each shock zone.

\subsubsection{Electron Synchrotron Emission}
\label{s4.4.1}

For a single power-law CRe distribution with the slope $q$, the synchrotron emissivity takes the power-law form, $j_{\rm syn}(\nu) \propto \nu^{-\alpha}$ with $\alpha = (q - 3)/2$.
With the volume-averaged CRe spectrum in Equation (\ref{fcreva}), on the other hand,
the synchrotron emissivity from each shock zone can be represented by the following broken power-law spectrum:
\begin{equation}
 j_{\rm syn}(\nu)\approx j_0 \cdot \left( \nu \over \nu_{\rm min} \right)^{-\alpha} \cdot \left(1 + \left(\zeta \over \zeta_0 \right)^b \right)^{-1/b}.
\end{equation}
Again, the fitting form, $(1 + (\zeta /\zeta_0)^b )^{-1/b}$, with the similarity variable, $\zeta \equiv \nu/\nu_{\rm br}$, along with the fitting parameters, $\zeta_0 \approx 0.095$ and $b \approx 0.65$,
is found from pertinent numerical calculations for the postshock advection and cooling.
Here, the ``break" frequency can be calculated from $p_{\rm br}$ in Equation (\ref{pbr}) and is given as
\begin{equation}
\nu_{\rm br} \approx 0.127~{\rm GHz}\left(t_{\rm adv} \over 10^8 {\rm yr}\right)^{-2} \left(B_{\rm e,2} \over 5\mu {\rm G}\right)^{-4}
\left({B_2\over \mu {\rm G}}\right).
\end{equation} 
The normalization factor $j_0$ at $\nu_{\rm min}$ depends on the CRe number density, the power-law slope $q$, and the postshock magnetic field strength $B_2$ as $j_0 \propto  \langle f_{\rm CRe}\rangle B_2^{(q-1)/2}$.

\subsubsection{Electron Inverse-Compton Emission}
\label{s4.4.2}

To compute the photon production rate of IC emission from each shock zone, we adopt the formula proposed in \citet{inoue1996} (see their equations [2.7]-[2.11]): 
\begin{equation}
q_{\rm IC}(\epsilon) = \int d\epsilon_0 n(\epsilon_0) \int d\gamma n_{\rm CRe}(\gamma) C(\epsilon, \gamma, \epsilon_0), 
\end{equation}
where $n(\epsilon_0)$ is the number density of CMB photons, which is equivalent to that of the black body radiation with the temperature $T = 2.757 {\rm K}(1 + z_r)$.
The volume-averaged CRe energy spectrum of each shock zone is $n_{\rm CRe}(\gamma) d\gamma = \langle f_{\rm CRe}(p)\rangle 4\pi p^2 dp$.
The function $C(\epsilon, \gamma, \epsilon_0)$ is the Compton kernel, where $\epsilon_0 m_e c^2$, $\epsilon m_e c^2$, and $\gamma m_e c^2$ are the energies of CMB photons, scattered photons, and CRe, respectively  \citep[see][]{jones1968}.

Then, the IC emissivity can be calculated as $j_{\rm IC}(\nu) = (h/4\pi) \epsilon q_{\rm IC}(\epsilon)$, where $\nu = \epsilon m_e c^2/h$ and
$h$ is the Planck constant. 
Below, we consider the electron IC emission in the two photon energy bands to be compared with observational data of existing facilities: 
$j_{\rm IC-X}$ in the HXR band of 20-80~keV (NuSTAR)
and $j_{\rm IC-\gamma}$ in the $\gamma$-ray band of 1-10~GeV (Fermi-LAT). 

\subsubsection{$\gamma$-ray Emission from $\pi^0$-decay}
\label{s4.4.3}

The $\gamma$-ray emission due to $\pi^0$-decay is computed by using the formula proposed in \citet{kelner2006} (see their equations [77]-[79]).
First, the production rate of pions with energy $E_{\pi}$ at each shock zone is calculated with
\begin{equation}
q_{\pi}(E_{\pi}) \approx  \frac{c n_2}{K_{\pi}}\sigma_{pp}\left(m_p c^2 + \frac{E_{\pi}}{K_{\pi}}\right) n_{\rm CRp}\left(m_p c^2 + \frac{E_{\pi}}{K_{\pi}}\right).
\end{equation}
Here, $K_{\pi} \approx 0.17$ is the fraction of the kinetic energy transferred from proton to pion, $ n_{\rm CRp}(E) dE =  f_{\rm CRp}(p) 4\pi p^2 dp$ is the CRp energy distribution at each shock zone, and $\sigma_{pp}(E)=(34.3 + 1.88 \phi + 0.25 \phi^2) \times [1 - (E_{\rm th}/E)^4]^2$ mb is the cross section, where $E_{\rm th} = 1.22$ GeV is the threshold energy of $p$-$p$ collision and $\phi = {\rm ln}(E/{\rm TeV})$. 
The production rate of $\gamma-$ray with energy $E_{\gamma}$ is given as
\begin{equation}
q_{\pi^0-\gamma}(E_{\gamma}) = 2\int_{E_{\rm min}}^{\infty} dE_{\pi} \frac{q_{\pi}(E_{\pi})}{\sqrt{E_{\pi}^2 - m_{\pi}^2 c^4}},
\end{equation}
where $E_{\rm min} = E_{\gamma} + m_{\pi}^2 c^4/(4E_{\gamma})$. 

Then, the $\gamma$-ray emissivity due to $\pi^0$-decay can be calculated as
$j_{\pi^0-\gamma}(\nu)=(h/4\pi)E_{\gamma} q_{\pi^0}$, where $\nu = E_{\gamma}/h$.
Below, we consider $j_{\pi^0-\gamma}(\nu)$ in the $\gamma$-ray band of 1-10~GeV (Fermi-LAT).

\subsection{Synthetic Multi-Wavelength Observations of Simulated Clusters}
\label{s4.5}

\begin{figure*}[t]
\vskip 0.0 cm
\hskip 0.0 cm
\centerline{\includegraphics[width=0.85\textwidth]{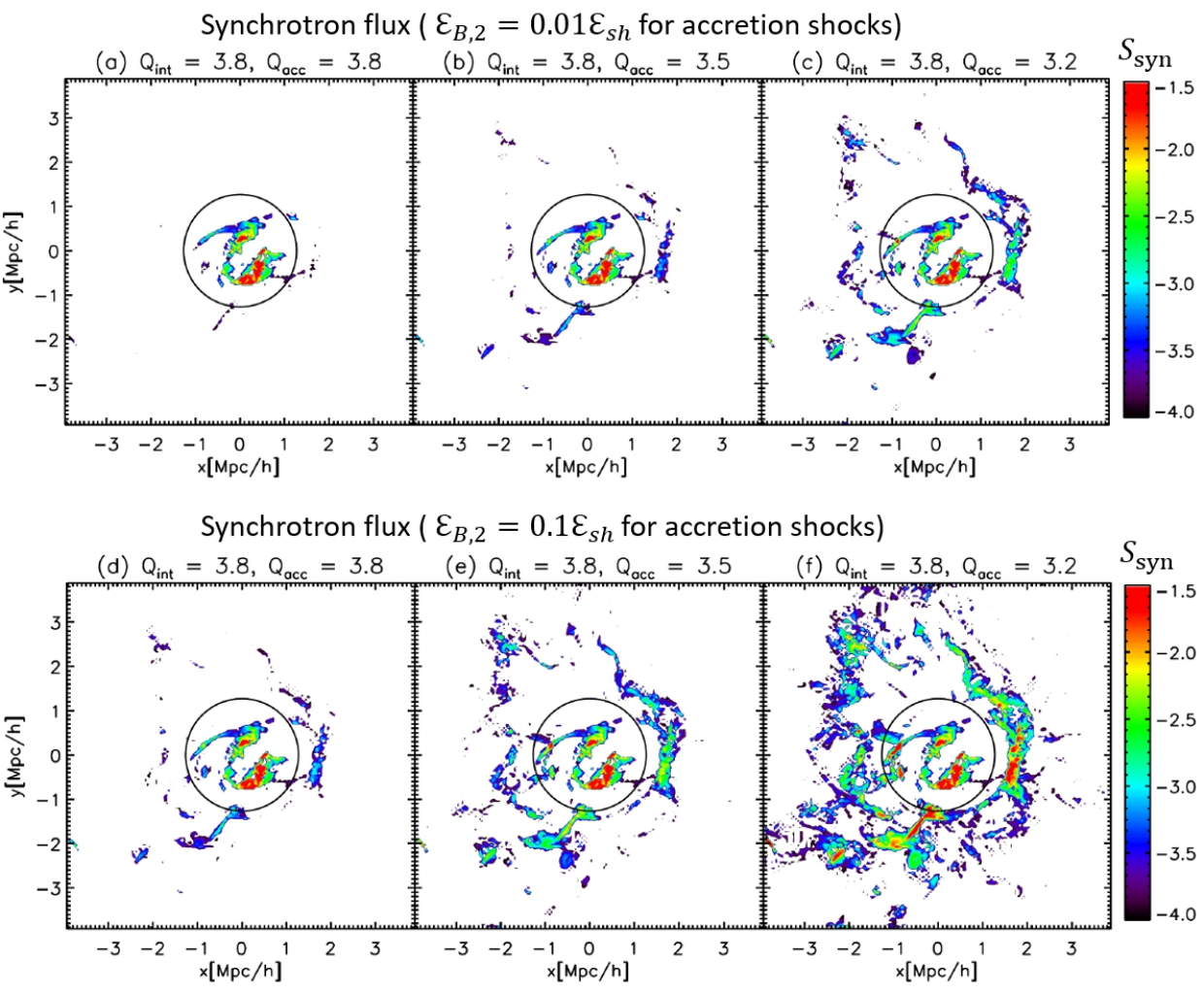}}
\vskip 0.1cm
\caption{Synthetic flux maps, $S_{\rm syn}(\nu)$, of the CL1 cluster, showing the synchrotron radiation at $\nu_{\rm obs} = 144$ MHz with a beam of $\theta^2 = 1' \times 1'$. 
The prescription for the magnetic field strength at accretion shocks, $\mathcal{E}_{B,2} = 0.01 \mathcal{E}_{\rm sh}$ is adopted in the top panels, while $\mathcal{E}_{B,2} = 0.1 \mathcal{E}_{\rm sh}$ in the bottom panels.
The values of the injection parameters, $Q_{\rm inj}$ and $Q_{\rm acc}$, are given at each panel.
Here, $S_{\rm syn}(\nu)$ is expressed in units of ${\rm Jy~beam^{-1}}$ and displayed in a logarithmic scale. 
A circle with the radius of $r = r_{200}$ is overlaid in each panel.
\label{f11}}
\end{figure*}

\begin{figure*}[t]
\vskip 0.0 cm
\hskip 0.0 cm
\centerline{\includegraphics[width=0.85\textwidth]{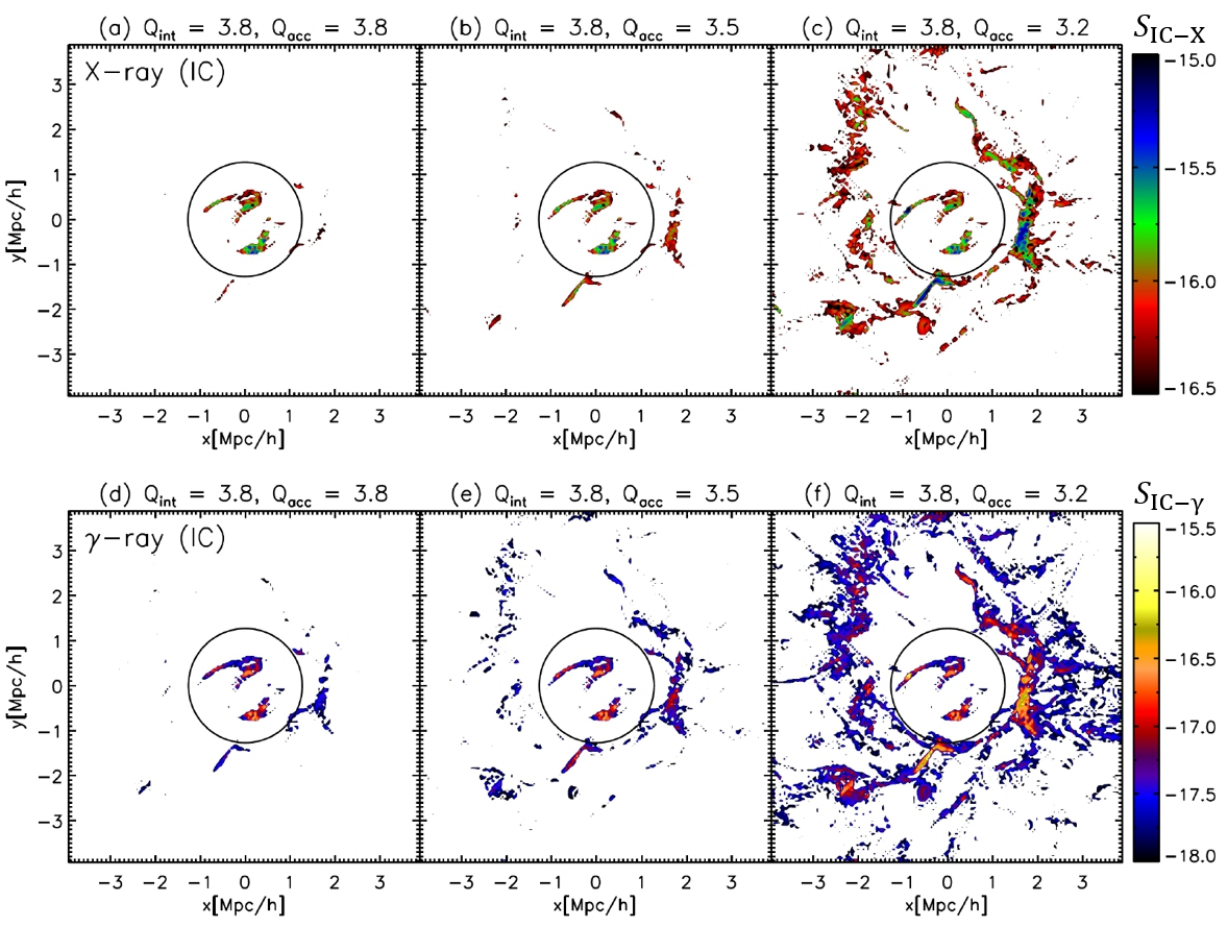}}
\vskip 0.1 cm
\caption{Synthetic flux maps of IC of the CL1 cluster. Panels (a)-(c): $S_{\rm IC-X}$ integrated in the energy range of 20-80 keV. Panels (d)-(f): $S_{\rm IC-\gamma}$ integrated in the energy range of 1-10 GeV. 
The adopted values of $Q_{\rm int}$ and $Q_{\rm acc}$ are given in each panel.
A pixel beam of $\theta^2 = 1' \times 1'$ is adopted uniformly for all the flux maps.
The expected fluxes are expressed in units of ${\rm erg~cm^{-2}~s^{-1}~arcmin^{-2}}$ and displayed in a logarithmic scale.
A circle with the radius of $r = r_{200}$ is overlaid in each panel.
\label{f12}}
\end{figure*}

As shown in Figure \ref{f10}, $\mathcal{N}_{\rm acc, CRp}$ and $\mathcal{N}_{\rm acc, CRe}$ at accretion shocks are similar in the three sample clusters, whereas $\mathcal{N}_{\rm int, CRp}$ and $\mathcal{N}_{\rm int, CRe}$ at internal shocks are the largest in CL1 among the three.
So we discuss nonthermal radiation from CL1 as an exemplary case. 

The relevant models, such as $Q_{\rm int}$ versus $Q_{\rm acc}$ and the $\beta=50$ versus $\epsilon_B$ prescriptions, are applied according to the types of shocks.
Both CRp and CRe are deposited to each numerical shock zone at $z_r=0$, without 
following their transport via advection, diffusion, and turbulent mixing throughout the structure formation history.
Then, nonthermal radiations are calculated with the one-zone approximation.
A numerical shock in our cosmological simulations is defined within a cubic grid zone of thickness $\Delta l = 48.8 h^{-1}$ kpc. 
The advection time for CRe to pass such thickness is $t_{\rm adv} = \Delta l/u_2 \approx 2.8 \times 10^8 {\rm yr} \cdot (\Delta l/48.8 h^{-1} {\rm kpc}) (u_2/250 {\rm km~s^{-1}})^{-1}$. 
Due to IC losses, for $\gamma \gtrsim 10^4$, $t_{\rm cool}$ is shorter than $t_{\rm adv}$, regardless of the magnetic field strength. 
This implies that high-energy CRe should have cooled down before exiting the shock zone, and hence the one-zone approximation would provide reasonable descriptions for nonthermal emission from CRe.
By contrast, CRp have negligible energy loss and accumulate in the ICM during the history of cluster formation. 
Hence, the estimations of $\pi^0$ $\gamma$-rays with the one-zone approximation should represent the radiation only around shocks, rather than over the whole cluster volume.

To construct synthetic flux maps, the intensity (surface brightness), $I(\nu)=\int_{los}j(\nu) dl$, is estimated by the integration along the line of sight and expressed in units of ${\rm erg~cm^{-2}~s^{-1}~Hz^{-1}~str^{-1}}$.
Without loss of generality, we choose the line of sight that is parallel to a coordinate axis, the $z$-axis, with the range of $[z-z_c]=[-4,+4]r_{200}$, where $z_c$ corresponds to the cluster center location. 
Then, the measured flux can be estimated approximately as 
$S(\nu)\approx I(\nu) \theta^2 (1+z_r)^{-3}$ for a pixel beam.
Here, we adopt the pixel beam of $\theta=1'$ to calculate $S_{\rm syn}$, $S_{\rm IC-X}$, and $S_{\rm IC-\gamma}$ in units of ${\rm erg~cm^{-2}~s^{-1}~Hz^{-1}~arcmin^{-2}}$.
To construct flux maps for specific observational facilities, such as LOFAR, NuSTAR, and Fermi LAT, the appropriate beam sizes should be used.
Considering the assumptions and limitations in our model calculations, however, we do not intend to make detailed comparisons between our synthetic maps and real observations of specific clusters.
Instead, we simply present the results with a fixed beam of $\theta=1'$.

Figure \ref{f11} displays the synchrotron flux, $S_{\rm syn}$, of the CL1 cluster with six different models: 
i.e., two values of $\epsilon_B$ for accretion shocks and three sets of $Q_{\rm int}$ and $Q_{\rm acc}$. 
A radio observation at $\nu_{\rm obs}=144$~MHz with a beam of $\theta^2 = 1' \times 1'$ is emulated to make a rough comparison with the observations of the Coma cluster using LOFAR  \citep[][]{bonafede2022}.  
The comparison of Figure \ref{f11} with Coma seems to imply the following.
(1) In the inner region of the sample cluster, the synchrotron emission estimated with $Q_{\rm int} = 3.8$ ranges as $S_{\rm syn}(\nu)\sim 10^{-3}-10^{-2}~{\rm Jy~beam^{-1}}$, which is in a reasonable agreement with the observations of Coma. 
(2) With $\epsilon_B = 0.01$ for accretion shocks, the model with $Q_{\rm acc} = 3.5$ reproduces reasonably well the observed flux levels in the outer region of Coma, $S_{\rm syn}(\nu)\sim 10^{-3}~{\rm Jy~beam^{-1}}$.
With $\epsilon_B = 0.1$, on the other hand, the model with $Q_{\rm acc} = 3.8$ seems to fit better the observations. 
(3) The models with an unrealistically small value, $Q_{\rm acc} = 3.2$, tend to overproduce the synchrotron emission in the outer region of the cluster.
As discussed in Sections \ref{s1} and \ref{s3}, PIC simulation studies appear to indicate that the thermal leakage injection recipe with $Q \sim 3.5-3.8$ could emulate reasonably the realistic spectra of shock-accelerated CRp and CRe.
So the results presented in Figure \ref{f11} are in line with such expectations.

The population of CRe that generates radio synchrotron emission shown in Figure \ref{f11} should yield diffuse HXR
 and $\gamma$-ray radiations through IC scattering of CMB photons.
So far the detection of IC HXR from clusters remains somewhat illusive, since only upper limits could be estimated with a statistical significance, partly owing to the limitations of multi-temperature modelings of thermal components \citep[e.g.,][]{cova2019,mirakhor2022}. 
Adopting $Q_{\rm int}=3.8$ and $Q_{\rm acc}=3.2-3.8$, we compute $S_{\rm IC-X}$ integrated for the NuSTAR 20-80~keV band, as shown in Figure \ref{f12}(a)-(c).
For internal shocks with $Q_{\rm int} = 3.8$, the flux level in the HXR band ranges $\sim 10^{-16}-10^{-15}~{\rm erg}~{\rm cm}^{-2}~s^{-1}~{\rm arcmin}^{-2}$ in the inner region.
The upper limit to non-thermal IC
emission is $5.1\times 10^{-12}{\rm erg}~{\rm cm}^{-2}~s^{-1}$ in the 20-80~keV band in a $12'\times 12'$ field of view of the Coma cluster \citep{gastaldello2015}.
Also \citet{cova2019} estimated an upper limit to the nonthermal radiation that comes from the central region of a $5'$ radius around Abell 523 ($T_{\rm X} \sim 5.3$ keV) as $3.2\times 10^{-14}~{\rm erg}~{\rm cm}^{-2}~s^{-1}$ in the 20-80~keV band.
Hence, our prediction for the HXR flux of the CL1 cluster is comparable to the upper limit for Abel 523, but lower than the upper limit for Coma.
By contrast, for accretion shocks with $Q_{\rm acc} = 3.8-3.5$, the 20-80~keV flux would be lower, $\lesssim 10^{-16}~{\rm erg}~{\rm cm}^{-2}~s^{-1}~{\rm arcmin}^{-2}$, in the outer region.
So we anticipate that detecting HXR emission from accretion shocks in the outer region of galaxy clusters would be quite challenging.
Note that the model with $Q_{\rm acc}=3.2$ shown in Figure \ref{f12}(c) predicts larger $S_{\rm IC-X}$, but is probably unrealistic, as we noted above.
 
In the inner region of typical clusters, diffuse $\gamma$-ray emission is expected to be produced mainly by $\pi^0$-decay due to high gas density
($j_{\pi^0 - \gamma} \propto n_2 n_{\rm CRp} \propto n_2^2$), whereas the electron IC $\gamma$-ray emission dominates in the outer region \citep[e.g.,][]{miniati2003,pinzke2010}.
So here we focus on the flux map of $S_{\rm IC-\gamma}$ for accretion shocks.
Figure \ref{f12}(d)-(f) show the IC $\gamma$-ray flux integrated for the Fermi-LAT 1-10~GeV band for the models
with $\epsilon_B = 0.01$ and three values of $Q_{\rm acc}$. 
In the model with $Q_{\rm acc} = 3.5$, for example, $S_{\rm IC-\gamma}\sim 10^{-18}-10^{-17} {\rm erg}~{\rm cm}^{-2}~{s}^{-1}~{\rm arcmin}^{-2}$ in the outer region. 

\begin{deluxetable}{ccccccccccccc}[t]
\tablecaption{Cluster-wide $\gamma$-ray Luminosity$^a$ \label{t3}}
\tabletypesize{\scriptsize}
\tablecolumns{12}
\tablenum{3}
\tablewidth{0pt}
\tablehead{
\colhead{Process} &
\colhead{$Q_{\rm int}$} &
\colhead{$Q_{\rm acc}$} &
\colhead{$L_{\gamma,{\rm int}}[10^{40} {\rm erg}~s^{-1}]$} &
\colhead{$L_{\gamma,{\rm acc}}[10^{40} {\rm erg}~s^{-1}]$}
}
\startdata
IC &  3.8 & 3.8 & 4.8 & 0.92\\
IC &  3.8 & 3.5 & 4.8 & 6.5\\
IC &  3.8 & 3.2 & 4.8 & 34\\
\hline
$\pi^0$-decay &  3.8 & 3.8 & 23 & 0.04\\
$\pi^0$-decay &  3.8 & 3.5 & 23 & 0.31\\
$\pi^0$-decay &  3.8 & 3.2 & 23 & 1.6\\
$\pi^0$-decay &  3.5 & 3.8 & 150 & 0.04\\
$\pi^0$-decay &  3.5 & 3.5 & 150 & 0.31\\
$\pi^0$-decay &  3.5 & 3.2 & 150 & 1.6\\
\enddata
\tablenotetext{a}{Estimated for the simulated CL1 cluster with $kT_{X} \approx 4.8$ keV and $r_{200} \approx 1.25 h^{-1}$ Mpc.}
\vspace{-0.2cm}
\end{deluxetable}

Since it would be unrealistic to construct the flux map of $\pi^0$ $\gamma$-ray emission without following properly the CRp transport,
we instead estimate in a somewhat crude way the ``cluster-wide" $\gamma$-ray luminosity in the 1-10~GeV band due to internal and accretion shocks as follows:
\begin{eqnarray}
L_{\gamma,\rm int}=\int_0^{r_{200}} \int_{\rm 1}^{\rm 10GeV} E_{\gamma} q_{\gamma} dE_{\gamma} dV, \nonumber \\ 
L_{\gamma,\rm acc}=\int_{r_{200}}^{4r_{200}}\int_{\rm 1}^{\rm 10GeV} E_{\gamma} q_{\gamma} dE_{\gamma} dV,
\end{eqnarray} 
where the photon production rates $q_{\gamma}$ due to electron IC and proton $\pi^0$-decay are considered separately.
Table \ref{t3} summarizes the results for different sets of $Q_{\rm int}$ and $Q_{\rm acc}$ for the two nonthermal processes.
In the cluster inner region the contribution from $\pi^0$-decay dominates over that from IC,
while it is the other way around in the outer region.
For the model with $Q_{\rm int}=3.8$ and $Q_{\rm acc}=3.5$, over the cluster-wise,
$L_{\pi^0-\gamma, \rm acc}/L_{\pi^0-\gamma, \rm int}\approx 0.01$ for $\pi^0$-decay emission,
while $L_{\rm IC-\gamma, \rm acc}/L_{\rm IC-\gamma, \rm int}\approx 1.4$ for IC emission.
In the cluster outer region, the IC contribution dominates over the $\pi^0$-decay contribution;
for the model with $Q_{\rm acc}=3.5$, $L_{\rm IC-\gamma, \rm acc}/L_{\pi^0-\gamma, \rm acc}\approx 20$.

In short, we find that our analytic DSA energy spectra for CRp and CRe described in Section \ref{s3} predict nonthermal emission from internal and accretion shocks, which does not contradict current multi-wavelength observations of galaxy clusters.

\section{Summary}
\label{s5}

The $\Lambda$CDM cosmological structure formation scenario predicts the formation of strong accretion shocks in the outer region of galaxy clusters, in addition to weak internal shocks in the ICM (see Table \ref{t1}).
Due to the extremely low gas density, the existence of putative accretion shocks has yet to be confirmed through the detection of nonthermal emission from CR particles produced via DSA.

To the best of our knowledge, the plasma physics of strong collisionless shocks in weakly-magnetized plasmas has not been investigated in details before.
In this work, we first showed that electrons can be pre-energized and injected to DSA at strong high-$\beta$ shocks, using
2D PIC simulations of perpendicular shocks with $M_{\rm s} = 10 - 10^2$, $T_1 = 10^4 - 10^6$~K, and $\beta = 10^2-10^3$. 
The electron preacceleration relies mainly on stochastic Fermi II acceleration owing to the ion-Weibel filamentation instability in the shock transition.
Although our PIC simulations were performed for a short kinetic timescale ($8\times 10^3 \omega_{\rm pi}^{-1}$), we expect that on sufficiently longer timescales suprathermal electrons would be injected to the full DSA process and accelerated to relativistic energies.

We then proposed analytic models for CRp and CRe energy spectra given in Equations (\ref{crp}) and (\ref{cre}), which can be employed for both weak internal shocks and strong accretion shocks, as long as the CRp acceleration remains
restricted within the test-particle limit.
They are basically rooted in the conventional thermal leakage injection model that have been explored in many previous studies \citep[e.g.][]{ryu2019,kang2020}; 
the power-law spectra emerge from the respective postshock Maxwellian distributions at $p_{\rm inj}=Qp_{\rm th,p}$ for CRp and at $p_{\rm min}=Qp_{\rm th,e}$ for CRe.
Assuming the same value of $Q$ for protons and electrons, the models yield the same injection fractions by number, $\xi_p=\xi_e$, as given in Equation (\ref{xipe}), and the CRp-to-CRe ratio, $K_{\rm p/e}$, as given in Equation (\ref{Kpe}).
Hence, this proposed DSA model is specified by a single injection parameter, $Q\sim 3.5-3.8$, which has been constrained reasonably well by many previous studies of collisionless shocks with a broad range of plasma parameters, as reviewed in the introduction.

We suggest that the proposed DSA spectra based on the thermal leakage injection model may provide ``generic'', yet physically-motivated, recipes for CR energy spectra to be employed for the studies of CR productions at cosmological shocks with a wide range of parameters.

To assess the suitability of the model recipes, we applied the proposed DSA spectra with $Q_{\rm int}\approx 3.5-3.8$ (for internal shocks) and $Q_{\rm acc}\approx 3.2-3.8$ (for accretion shocks) to identified ``shock zones'' in and around three simulated clusters drawn from cosmological hydrodynamic simulations of large-scale structure formation.
We then estimated nonthermal emissions due to CRp and CRe, including radio synchrotron, IC emission in HXR and $\gamma$-ray bands, and $\pi^0$-decay $\gamma$-ray emission.
We obtained the synthetic radio maps due to internal and accretion shocks of a simulated cluster CL1 ($kT_X \approx 4.8$~keV) with $Q_{\rm int}=3.8$ and $Q_{\rm acc}=3.5-3.8$ that are compatible with the recent LOFAR observations of the Coma cluster \citep[][]{bonafede2022}.
In the 20-80 keV HXR band, the predicted flux levels in the inner region of CL1 tend to lie below the upper limits for several clusters observed by NuSTAR \citep[e.g.,][]{gastaldello2015, cova2019,mirakhor2022}, whereas the HXR flux from accretion shocks in the outer region would be orders of magnitude smaller.

We also calculated the cluster-wide $\gamma$-ray luminosity, $L_{\gamma}$, in the 1-10 GeV band due to IC and $\pi^0$-decay emission. 
Comparison of the inner and outer regions indicates that $L_{\pi^0-\gamma,\rm int}\gg L_{\pi^0-\gamma,\rm acc}$ for $\pi^0$ $\gamma$-ray emission (see Table \ref{t3}), mainly because the gas density is much higher in the inner region.
In contrast, the IC contribution dominates over the $\pi^0$-decay contribution in the cluster outer region, $L_{\rm IC-\gamma,\rm acc}\gg L_{\pi^0-\gamma,\rm acc}$,  which is consistent with the previous findings \citep[e.g.,][]{miniati2003, keshet2003, pinzke2010}.

\begin{acknowledgments}
The authors thank the anonymous referee for constructive comments.
This work was supported by the National Research Foundation (NRF) of Korea through grants 2016R1A5A1013277, 2020R1A2C2102800, 2020R1F1A1048189, and  RS-2022-00197685.  
Some of simulations were performed using the high performance computing resources of the UNIST Supercomputing Center.

\end{acknowledgments}

\bibliography{ref_shock}{}

\begin{thebibliography}{}
\expandafter\ifx\csname natexlab\endcsname\relax\def\natexlab#1{#1}\fi
\providecommand{\url}[1]{\href{#1}{#1}}
\providecommand{\dodoi}[1]{doi:~\href{http://doi.org/#1}{\nolinkurl{#1}}}
\providecommand{\doeprint}[1]{\href{http://ascl.net/#1}{\nolinkurl{http://ascl.net/#1}}}
\providecommand{\doarXiv}[1]{\href{https://arxiv.org/abs/#1}{\nolinkurl{https://arxiv.org/abs/#1}}}

\bibitem[{{Ackermann} {et~al.}(2016){Ackermann}, {Ajello}, {Albert}, {Atwood},
  {Baldini}, {Ballet}, {Barbiellini}, {Bastieri}, {Bechtol}, {Bellazzini},
  {Bissaldi}, {Blandford}, {Bloom}, {Bonino}, {Bottacini}, {Bregeon}, {Bruel},
  {Buehler}, {Caliandro}, {Cameron}, {Caragiulo}, {Caraveo}, {Casandjian},
  {Cavazzuti}, {Cecchi}, {Charles}, {Chekhtman}, {Chiaro}, {Ciprini},
  {Cohen-Tanugi}, {Conrad}, {Cutini}, {D'Ammando}, {de Angelis}, {de Palma},
  {Desiante}, {Digel}, {Di Venere}, {Drell}, {Favuzzi}, {Fegan}, {Fukazawa},
  {Funk}, {Fusco}, {Gargano}, {Gasparrini}, {Giglietto}, {Giordano},
  {Giroletti}, {Godfrey}, {Green}, {Grenier}, {Guiriec}, {Hays}, {Hewitt},
  {Horan}, {J{\'o}hannesson}, {Kuss}, {Larsson}, {Latronico}, {Li}, {Li},
  {Longo}, {Loparco}, {Lovellette}, {Lubrano}, {Madejski}, {Maldera},
  {Manfreda}, {Mayer}, {Mazziotta}, {Michelson}, {Mitthumsiri}, {Mizuno},
  {Monzani}, {Morselli}, {Moskalenko}, {Murgia}, {Nuss}, {Ohsugi}, {Orienti},
  {Orlando}, {Ormes}, {Paneque}, {Pesce-Rollins}, {Petrosian}, {Piron},
  {Pivato}, {Porter}, {Rain{\`o}}, {Rando}, {Razzano}, {Reimer}, {Reimer},
  {S{\'a}nchez-Conde}, {Sgr{\`o}}, {Siskind}, {Spada}, {Spandre}, {Spinelli},
  {Tajima}, {Takahashi}, {Thayer}, {Tibaldo}, {Torres}, {Tosti}, {Troja},
  {Vianello}, {Wood}, {Zimmer}, {Fermi-LAT Collaboration}, \&
  {Rephaeli}}]{ackermann2016}
{Ackermann}, M., {Ajello}, M., {Albert}, A., {et~al.} 2016, \apj, 819, 149,
  \dodoi{10.3847/0004-637X/819/2/149}

\bibitem[{{Adam} {et~al.}(2021){Adam}, {Goksu}, {Brown}, {Rudnick}, \&
  {Ferrari}}]{adam2021}
{Adam}, R., {Goksu}, H., {Brown}, S., {Rudnick}, L., \& {Ferrari}, C. 2021,
  \aap, 648, A60, \dodoi{10.1051/0004-6361/202039660}

\bibitem[{{Amano} \& {Hoshino}(2009)}]{amano2009}
{Amano}, T., \& {Hoshino}, M. 2009, \apj, 690, 244,
  \dodoi{10.1088/0004-637X/690/1/244}

\bibitem[{{Amano} {et~al.}(2022){Amano}, {Matsumoto}, {Bohdan}, {Kobzar},
  {Matsukiyo}, {Oka}, {Niemiec}, {Pohl}, \& {Hoshino}}]{amano2022}
{Amano}, T., {Matsumoto}, Y., {Bohdan}, A., {et~al.} 2022, Reviews of Modern
  Plasma Physics, 6, 29, \dodoi{10.1007/s41614-022-00093-1}

\bibitem[{Arbutina \& Zeković(2021)}]{arbutina2021}
Arbutina, B., \& Zeković, V. 2021, Astroparticle Physics, 127, 102546,
  \dodoi{https://doi.org/10.1016/j.astropartphys.2020.102546}

\bibitem[{{Balogh} \& {Treumann}(2013)}]{balogh2013}
{Balogh}, A., \& {Treumann}, R.~A. 2013, {Physics of Collisionless Shocks},
  Vol.~12, \dodoi{10.1007/978-1-4614-6099-2}

\bibitem[{{Bell}(1978)}]{bell1978}
{Bell}, A.~R. 1978, \mnras, 182, 147, \dodoi{10.1093/mnras/182.2.147}

\bibitem[{{Bell}(2004)}]{bell2004}
---. 2004, \mnras, 353, 550, \dodoi{10.1111/j.1365-2966.2004.08097.x}

\bibitem[{{Blandford} \& {Ostriker}(1978)}]{blandford1978}
{Blandford}, R.~D., \& {Ostriker}, J.~P. 1978, \apjl, 221, L29,
  \dodoi{10.1086/182658}

\bibitem[{{Bohdan} {et~al.}(2017){Bohdan}, {Niemiec}, {Kobzar}, \&
  {Pohl}}]{bohdan2017}
{Bohdan}, A., {Niemiec}, J., {Kobzar}, O., \& {Pohl}, M. 2017, \apj, 847, 71,
  \dodoi{10.3847/1538-4357/aa872a}

\bibitem[{{Bohdan} {et~al.}(2019{\natexlab{a}}){Bohdan}, {Niemiec}, {Pohl},
  {Matsumoto}, {Amano}, \& {Hoshino}}]{bohdan2019a}
{Bohdan}, A., {Niemiec}, J., {Pohl}, M., {et~al.} 2019{\natexlab{a}}, \apj,
  878, 5, \dodoi{10.3847/1538-4357/ab1b6d}

\bibitem[{{Bohdan} {et~al.}(2019{\natexlab{b}}){Bohdan}, {Niemiec}, {Pohl},
  {Matsumoto}, {Amano}, \& {Hoshino}}]{bohdan2019b}
---. 2019{\natexlab{b}}, \apj, 885, 10, \dodoi{10.3847/1538-4357/ab43cf}

\bibitem[{{Bohdan} {et~al.}(2021){Bohdan}, {Pohl}, {Niemiec}, {Morris},
  {Matsumoto}, {Amano}, {Hoshino}, \& {Sulaiman}}]{bohdan2021}
{Bohdan}, A., {Pohl}, M., {Niemiec}, J., {et~al.} 2021, \prl, 126, 095101,
  \dodoi{10.1103/PhysRevLett.126.095101}

\bibitem[{{Bohdan} {et~al.}(2020){Bohdan}, {Pohl}, {Niemiec}, {Vafin},
  {Matsumoto}, {Amano}, \& {Hoshino}}]{bohdan2020a}
---. 2020, \apj, 893, 6, \dodoi{10.3847/1538-4357/ab7cd6}

\bibitem[{Bonafede {et~al.}(2022)Bonafede, Brunetti, Rudnick, Vazza, Bourdin,
  Giovannini, Shimwell, Zhang, Mazzotta, Simionescu, Biava, Bonnassieux,
  Brienza, Brüggen, Rajpurohit, Riseley, Stuardi, Feretti, Tasse, Botteon,
  Carretti, Cassano, Cuciti, de~Gasperin, Gastaldello, Rossetti, Rottgering,
  Venturi, \& van Weeren}]{bonafede2022}
Bonafede, A., Brunetti, G., Rudnick, L., {et~al.} 2022,
  \dodoi{10.48550/ARXIV.2203.01958}

\bibitem[{Buneman(1993)}]{buneman1993}
Buneman, O. 1993, in Computer Space Plasma Physics: Simulation Techniques and
  Software, ed. H. Matsumoto \& Y. Omura, 67

\bibitem[{{Caprioli} {et~al.}(2010){Caprioli}, {Amato}, \&
  {Blasi}}]{caprioli2010}
{Caprioli}, D., {Amato}, E., \& {Blasi}, P. 2010, Astroparticle Physics, 33,
  307, \dodoi{10.1016/j.astropartphys.2010.03.001}

\bibitem[{{Caprioli} \& {Spitkovsky}(2014{\natexlab{a}})}]{caprioli2014a}
{Caprioli}, D., \& {Spitkovsky}, A. 2014{\natexlab{a}}, \apj, 783, 91,
  \dodoi{10.1088/0004-637X/783/2/91}

\bibitem[{{Caprioli} \& {Spitkovsky}(2014{\natexlab{b}})}]{caprioli2014b}
---. 2014{\natexlab{b}}, \apj, 794, 46, \dodoi{10.1088/0004-637X/794/1/46}

\bibitem[{{Carilli} {et~al.}(1991){Carilli}, {Perley}, {Dreher}, \&
  {Leahy}}]{carilli1991}
{Carilli}, C.~L., {Perley}, R.~A., {Dreher}, J.~W., \& {Leahy}, J.~P. 1991,
  \apj, 383, 554, \dodoi{10.1086/170813}

\bibitem[{{Cova} {et~al.}(2019){Cova}, {Gastaldello}, {Wik}, {Boschin},
  {Botteon}, {Brunetti}, {Buote}, {De Grandi}, {Eckert}, {Ettori}, {Feretti},
  {Gaspari}, {Ghizzardi}, {Giovannini}, {Girardi}, {Govoni}, {Molendi},
  {Murgia}, {Rossetti}, \& {Vacca}}]{cova2019}
{Cova}, F., {Gastaldello}, F., {Wik}, D.~R., {et~al.} 2019, \aap, 628, A83,
  \dodoi{10.1051/0004-6361/201834644}

\bibitem[{{Drury}(1983)}]{drury1983}
{Drury}, L.~O. 1983, Reports on Progress in Physics, 46, 973,
  \dodoi{10.1088/0034-4885/46/8/002}

\bibitem[{{En{\ss}lin} {et~al.}(1999){En{\ss}lin}, {Lieu}, \&
  {Biermann}}]{ensslin1999}
{En{\ss}lin}, T.~A., {Lieu}, R., \& {Biermann}, P.~L. 1999, \aap, 344, 409.
\newblock \doarXiv{astro-ph/9808139}

\bibitem[{{Fiuza} {et~al.}(2012){Fiuza}, {Fonseca}, {Tonge}, {Mori}, \&
  {Silva}}]{fiuza2012}
{Fiuza}, F., {Fonseca}, R.~A., {Tonge}, J., {Mori}, W.~B., \& {Silva}, L.~O.
  2012, \prl, 108, 235004, \dodoi{10.1103/PhysRevLett.108.235004}

\bibitem[{Fiuza {et~al.}(2020)Fiuza, Swadling, Grassi, Rinderknecht, Higginson,
  Ryutov, Bruulsema, Drake, Funk, Glenzer, {et~al.}}]{fiuza2020}
Fiuza, F., Swadling, G., Grassi, A., {et~al.} 2020, Nature physics, 16, 916

\bibitem[{Fried(1959)}]{fried1959}
Fried, B.~D. 1959, The Physics of Fluids, 2, 337, \dodoi{10.1063/1.1705933}

\bibitem[{{Gastaldello} {et~al.}(2015){Gastaldello}, {Wik}, {Molendi},
  {Westergaard}, {Hornstrup}, {Madejski}, {Ferreira}, {Boggs}, {Christensen},
  {Craig}, {Grefenstette}, {Hailey}, {Harrison}, {Madsen}, {Stern}, \&
  {Zhang}}]{gastaldello2015}
{Gastaldello}, F., {Wik}, D.~R., {Molendi}, S., {et~al.} 2015, \apj, 800, 139,
  \dodoi{10.1088/0004-637X/800/2/139}

\bibitem[{{Giacalone} \& {Jokipii}(2007)}]{giacalone2007}
{Giacalone}, J., \& {Jokipii}, J.~R. 2007, \apjl, 663, L41,
  \dodoi{10.1086/519994}

\bibitem[{{Gieseler} {et~al.}(2000){Gieseler}, {Jones}, \&
  {Kang}}]{gieseler2000}
{Gieseler}, U.~D.~J., {Jones}, T.~W., \& {Kang}, H. 2000, \aap, 364, 911.
\newblock \doarXiv{astro-ph/0011058}

\bibitem[{{Guo} {et~al.}(2014{\natexlab{a}}){Guo}, {Sironi}, \&
  {Narayan}}]{guo2014a}
{Guo}, X., {Sironi}, L., \& {Narayan}, R. 2014{\natexlab{a}}, \apj, 794, 153,
  \dodoi{10.1088/0004-637X/794/2/153}

\bibitem[{{Guo} {et~al.}(2014{\natexlab{b}}){Guo}, {Sironi}, \&
  {Narayan}}]{guo2014b}
---. 2014{\natexlab{b}}, \apj, 797, 47, \dodoi{10.1088/0004-637X/797/1/47}

\bibitem[{Ha {et~al.}(2021)Ha, Kim, Ryu, \& Kang}]{ha2021}
Ha, J.-H., Kim, S., Ryu, D., \& Kang, H. 2021, \apj, 915, 18,
  \dodoi{10.3847/1538-4357/abfb68}

\bibitem[{{Ha} {et~al.}(2018{\natexlab{a}}){Ha}, {Ryu}, \& {Kang}}]{ha2018a}
{Ha}, J.-H., {Ryu}, D., \& {Kang}, H. 2018{\natexlab{a}}, \apj, 857, 26,
  \dodoi{10.3847/1538-4357/aab4a2}

\bibitem[{{Ha} {et~al.}(2018{\natexlab{b}}){Ha}, {Ryu}, {Kang}, \& {van
  Marle}}]{ha2018b}
{Ha}, J.-H., {Ryu}, D., {Kang}, H., \& {van Marle}, A.~J. 2018{\natexlab{b}},
  \apj, 864, 105, \dodoi{10.3847/1538-4357/aad634}

\bibitem[{{Hoeft} {et~al.}(2008){Hoeft}, {Br{\"u}ggen}, {Yepes},
  {Gottl{\"o}ber}, \& {Schwope}}]{hoeft2008}
{Hoeft}, M., {Br{\"u}ggen}, M., {Yepes}, G., {Gottl{\"o}ber}, S., \& {Schwope},
  A. 2008, \mnras, 391, 1511, \dodoi{10.1111/j.1365-2966.2008.13955.x}

\bibitem[{{Hong} {et~al.}(2014){Hong}, {Ryu}, {Kang}, \& {Cen}}]{hong2014}
{Hong}, S.~E., {Ryu}, D., {Kang}, H., \& {Cen}, R. 2014, \apj, 785, 133,
  \dodoi{10.1088/0004-637X/785/2/133}

\bibitem[{{Huntington} {et~al.}(2015){Huntington}, {Fiuza}, {Ross}, {Zylstra},
  {Drake}, {Froula}, {Gregori}, {Kugland}, {Kuranz}, {Levy}, {Li}, {Meinecke},
  {Morita}, {Petrasso}, {Plechaty}, {Remington}, {Ryutov}, {Sakawa},
  {Spitkovsky}, {Takabe}, \& {Park}}]{huntington2015}
{Huntington}, C.~M., {Fiuza}, F., {Ross}, J.~S., {et~al.} 2015, Nature Physics,
  11, 173, \dodoi{10.1038/nphys3178}

\bibitem[{{Inoue} \& {Takahara}(1996)}]{inoue1996}
{Inoue}, S., \& {Takahara}, F. 1996, \apj, 463, 555, \dodoi{10.1086/177270}

\bibitem[{{Ji} {et~al.}(2016){Ji}, {Oh}, {Ruszkowski}, \&
  {Markevitch}}]{ji2016}
{Ji}, S., {Oh}, S.~P., {Ruszkowski}, M., \& {Markevitch}, M. 2016, \mnras, 463,
  3989, \dodoi{10.1093/mnras/stw2320}

\bibitem[{Jones(1968)}]{jones1968}
Jones, F.~C. 1968, Phys. Rev., 167, 1159, \dodoi{10.1103/PhysRev.167.1159}

\bibitem[{{Kang}(2020)}]{kang2020}
{Kang}, H. 2020, Journal of Korean Astronomical Society, 53, 59,
  \dodoi{10.5303/JKAS.2020.53.3.59}

\bibitem[{{Kang} {et~al.}(2002){Kang}, {Jones}, \& {Gieseler}}]{kang2002}
{Kang}, H., {Jones}, T.~W., \& {Gieseler}, U.~D.~J. 2002, \apj, 579, 337,
  \dodoi{10.1086/342724}

\bibitem[{{Kang} \& {Ryu}(2010)}]{kang2010}
{Kang}, H., \& {Ryu}, D. 2010, \apj, 721, 886,
  \dodoi{10.1088/0004-637X/721/1/886}

\bibitem[{{Kang} \& {Ryu}(2013)}]{kang2013}
---. 2013, \apj, 764, 95, \dodoi{10.1088/0004-637X/764/1/95}

\bibitem[{{Kang} {et~al.}(2007){Kang}, {Ryu}, {Cen}, \& {Ostriker}}]{kang2007}
{Kang}, H., {Ryu}, D., {Cen}, R., \& {Ostriker}, J.~P. 2007, \apj, 669, 729,
  \dodoi{10.1086/521717}

\bibitem[{Kang {et~al.}(2019)Kang, Ryu, \& Ha}]{kang2019}
Kang, H., Ryu, D., \& Ha, J.-H. 2019, The Astrophysical Journal, 876, 79,
  \dodoi{10.3847/1538-4357/ab16d1}

\bibitem[{{Kang} {et~al.}(1996){Kang}, {Ryu}, \& {Jones}}]{kang1996}
{Kang}, H., {Ryu}, D., \& {Jones}, T.~W. 1996, \apj, 456, 422,
  \dodoi{10.1086/176666}

\bibitem[{{Kato} \& {Takabe}(2008)}]{kato2008}
{Kato}, T.~N., \& {Takabe}, H. 2008, \apjl, 681, L93, \dodoi{10.1086/590387}

\bibitem[{{Kato} \& {Takabe}(2010)}]{kato2010}
---. 2010, \apj, 721, 828, \dodoi{10.1088/0004-637X/721/1/828}

\bibitem[{{Katou} \& {Amano}(2019)}]{katou2019}
{Katou}, T., \& {Amano}, T. 2019, \apj, 874, 119,
  \dodoi{10.3847/1538-4357/ab0d8a}

\bibitem[{{Kelner} {et~al.}(2006){Kelner}, {Aharonian}, \&
  {Bugayov}}]{kelner2006}
{Kelner}, S.~R., {Aharonian}, F.~A., \& {Bugayov}, V.~V. 2006, \prd, 74,
  034018, \dodoi{10.1103/PhysRevD.74.034018}

\bibitem[{{Keshet} {et~al.}(2017){Keshet}, {Kushnir}, {Loeb}, \&
  {Waxman}}]{keshet2017}
{Keshet}, U., {Kushnir}, D., {Loeb}, A., \& {Waxman}, E. 2017, \apj, 845, 24,
  \dodoi{10.3847/1538-4357/aa794b}

\bibitem[{{Keshet} {et~al.}(2003){Keshet}, {Waxman}, {Loeb}, {Springel}, \&
  {Hernquist}}]{keshet2003}
{Keshet}, U., {Waxman}, E., {Loeb}, A., {Springel}, V., \& {Hernquist}, L.
  2003, \apj, 585, 128, \dodoi{10.1086/345946}

\bibitem[{{Kim} {et~al.}(2020){Kim}, {Ha}, {Ryu}, \& {Kang}}]{kim2020}
{Kim}, S., {Ha}, J.-H., {Ryu}, D., \& {Kang}, H. 2020, \apj, 892, 85,
  \dodoi{10.3847/1538-4357/ab7cd9}

\bibitem[{{Kim} {et~al.}(2021){Kim}, {Ha}, {Ryu}, \& {Kang}}]{kim2021}
---. 2021, \apj, 913, 35, \dodoi{10.3847/1538-4357/abf1e1}

\bibitem[{{Kobzar} {et~al.}(2021){Kobzar}, {Niemiec}, {Amano}, {Hoshino},
  {Matsukiyo}, {Matsumoto}, \& {Pohl}}]{kobzar2021}
{Kobzar}, O., {Niemiec}, J., {Amano}, T., {et~al.} 2021, \apj, 919, 97,
  \dodoi{10.3847/1538-4357/ac1107}

\bibitem[{{Komatsu} {et~al.}(2011){Komatsu}, {Smith}, {Dunkley}, {Bennett},
  {Gold}, {Hinshaw}, {Jarosik}, {Larson}, {Nolta}, {Page}, {Spergel},
  {Halpern}, {Hill}, {Kogut}, {Limon}, {Meyer}, {Odegard}, {Tucker}, {Weiland},
  {Wollack}, \& {Wright}}]{komatsu2011}
{Komatsu}, E., {Smith}, K.~M., {Dunkley}, J., {et~al.} 2011, \apjs, 192, 18,
  \dodoi{10.1088/0067-0049/192/2/18}

\bibitem[{{Kushnir} \& {Waxman}(2010)}]{kushnir2010}
{Kushnir}, D., \& {Waxman}, E. 2010, \jcap, 2010, 025,
  \dodoi{10.1088/1475-7516/2010/02/025}

\bibitem[{{Loeb} \& {Waxman}(2000)}]{loeb2000}
{Loeb}, A., \& {Waxman}, E. 2000, \nat, 405, 156, \dodoi{10.1038/35012018}

\bibitem[{{Malkov} \& {Drury}(2001)}]{malkov2001}
{Malkov}, M.~A., \& {Drury}, L.~O. 2001, Reports on Progress in Physics, 64,
  429, \dodoi{10.1088/0034-4885/64/4/201}

\bibitem[{{Malkov} \& {V{\"o}lk}(1998)}]{malkov1998}
{Malkov}, M.~A., \& {V{\"o}lk}, H.~J. 1998, Advances in Space Research, 21,
  551, \dodoi{10.1016/S0273-1177(97)00961-7}

\bibitem[{{Marcowith} {et~al.}(2016){Marcowith}, {Bret}, {Bykov}, {Dieckman},
  {O'C Drury}, {Lemb{\`e}ge}, {Lemoine}, {Morlino}, {Murphy}, {Pelletier},
  {Plotnikov}, {Reville}, {Riquelme}, {Sironi}, \& {Stockem
  Novo}}]{marcowith2016}
{Marcowith}, A., {Bret}, A., {Bykov}, A., {et~al.} 2016, Reports on Progress in
  Physics, 79, 046901, \dodoi{10.1088/0034-4885/79/4/046901}

\bibitem[{{Matsumoto} {et~al.}(2012){Matsumoto}, {Amano}, \&
  {Hoshino}}]{matsumoto2012}
{Matsumoto}, Y., {Amano}, T., \& {Hoshino}, M. 2012, in AGU Fall Meeting
  Abstracts, Vol. 2012, SH33F--03

\bibitem[{{Matsumoto} {et~al.}(2015){Matsumoto}, {Amano}, {Kato}, \&
  {Hoshino}}]{matsumoto2015}
{Matsumoto}, Y., {Amano}, T., {Kato}, T.~N., \& {Hoshino}, M. 2015, Science,
  347, 974, \dodoi{10.1126/science.1260168}

\bibitem[{Matsumoto {et~al.}(2017)Matsumoto, Amano, Kato, \&
  Hoshino}]{matsumoto2017}
Matsumoto, Y., Amano, T., Kato, T.~N., \& Hoshino, M. 2017, Phys. Rev. Lett.,
  119, 105101, \dodoi{10.1103/PhysRevLett.119.105101}

\bibitem[{{Miniati}(2003)}]{miniati2003}
{Miniati}, F. 2003, \mnras, 342, 1009, \dodoi{10.1046/j.1365-8711.2003.06647.x}

\bibitem[{{Miniati} {et~al.}(2000){Miniati}, {Ryu}, {Kang}, {Jones}, {Cen}, \&
  {Ostriker}}]{miniati2000}
{Miniati}, F., {Ryu}, D., {Kang}, H., {et~al.} 2000, \apj, 542, 608,
  \dodoi{10.1086/317027}

\bibitem[{{Mirakhor} {et~al.}(2022){Mirakhor}, {Walker}, {Runge}, \&
  {Diwanji}}]{mirakhor2022}
{Mirakhor}, M.~S., {Walker}, S.~A., {Runge}, J., \& {Diwanji}, P. 2022, \mnras,
  516, 1855, \dodoi{10.1093/mnras/stac2379}

\bibitem[{{Niemiec} {et~al.}(2019){Niemiec}, {Kobzar}, {Amano}, {Hoshino},
  {Matsukiyo}, {Matsumoto}, \& {Pohl}}]{niemiec2019}
{Niemiec}, J., {Kobzar}, O., {Amano}, T., {et~al.} 2019, in International
  Cosmic Ray Conference, Vol.~36, 36th International Cosmic Ray Conference
  (ICRC2019), 368

\bibitem[{{Park} {et~al.}(2015){Park}, {Caprioli}, \& {Spitkovsky}}]{park2015}
{Park}, J., {Caprioli}, D., \& {Spitkovsky}, A. 2015, \prl, 114, 085003,
  \dodoi{10.1103/PhysRevLett.114.085003}

\bibitem[{{Pfrommer} {et~al.}(2006){Pfrommer}, {Springel}, {En{\ss}lin}, \&
  {Jubelgas}}]{pfrommer2006}
{Pfrommer}, C., {Springel}, V., {En{\ss}lin}, T.~A., \& {Jubelgas}, M. 2006,
  \mnras, 367, 113, \dodoi{10.1111/j.1365-2966.2005.09953.x}

\bibitem[{{Pinzke} \& {Pfrommer}(2010)}]{pinzke2010}
{Pinzke}, A., \& {Pfrommer}, C. 2010, \mnras, 409, 449,
  \dodoi{10.1111/j.1365-2966.2010.17328.x}

\bibitem[{{Riquelme} \& {Spitkovsky}(2011)}]{riquelme2011}
{Riquelme}, M.~A., \& {Spitkovsky}, A. 2011, \apj, 733, 63,
  \dodoi{10.1088/0004-637X/733/1/63}

\bibitem[{{Ryu} {et~al.}(2008){Ryu}, {Kang}, {Cho}, \& {Das}}]{ryu2008}
{Ryu}, D., {Kang}, H., {Cho}, J., \& {Das}, S. 2008, Science, 320, 909,
  \dodoi{10.1126/science.1154923}

\bibitem[{{Ryu} {et~al.}(2019){Ryu}, {Kang}, \& {Ha}}]{ryu2019}
{Ryu}, D., {Kang}, H., \& {Ha}, J.-H. 2019, \apj, 883, 60,
  \dodoi{10.3847/1538-4357/ab3a3a}

\bibitem[{{Ryu} {et~al.}(2003){Ryu}, {Kang}, {Hallman}, \& {Jones}}]{ryu2003}
{Ryu}, D., {Kang}, H., {Hallman}, E., \& {Jones}, T.~W. 2003, \apj, 593, 599,
  \dodoi{10.1086/376723}

\bibitem[{{Ryu} {et~al.}(1993){Ryu}, {Ostriker}, {Kang}, \& {Cen}}]{ryu1993}
{Ryu}, D., {Ostriker}, J.~P., {Kang}, H., \& {Cen}, R. 1993, \apj, 414, 1,
  \dodoi{10.1086/173051}

\bibitem[{{Schaal} \& {Springel}(2015)}]{schaal2015}
{Schaal}, K., \& {Springel}, V. 2015, \mnras, 446, 3992,
  \dodoi{10.1093/mnras/stu2386}

\bibitem[{{Scharf} \& {Mukherjee}(2002)}]{scharf2002}
{Scharf}, C.~A., \& {Mukherjee}, R. 2002, \apj, 580, 154,
  \dodoi{10.1086/343035}

\bibitem[{{Schlickeiser} \& {Shukla}(2003)}]{schlickeiser2003}
{Schlickeiser}, R., \& {Shukla}, P.~K. 2003, \apjl, 599, L57,
  \dodoi{10.1086/381246}

\bibitem[{{Sironi} \& {Spitkovsky}(2009)}]{sironi2009}
{Sironi}, L., \& {Spitkovsky}, A. 2009, \apjl, 707, L92,
  \dodoi{10.1088/0004-637X/707/1/L92}

\bibitem[{{Sironi} {et~al.}(2013){Sironi}, {Spitkovsky}, \&
  {Arons}}]{sironi2013}
{Sironi}, L., {Spitkovsky}, A., \& {Arons}, J. 2013, \apj, 771, 54,
  \dodoi{10.1088/0004-637X/771/1/54}

\bibitem[{{Skillman} {et~al.}(2008){Skillman}, {O'Shea}, {Hallman}, {Burns}, \&
  {Norman}}]{skillman2008}
{Skillman}, S.~W., {O'Shea}, B.~W., {Hallman}, E.~J., {Burns}, J.~O., \&
  {Norman}, M.~L. 2008, \apj, 689, 1063, \dodoi{10.1086/592496}

\bibitem[{Spitkovsky(2005)}]{spitkovsky2005}
Spitkovsky, A. 2005, in AIP Conf. Proc., Astrophysical Sources of High Energy
  Particles and Radiation, ed. T. Bulik, B. Rudak, \& G. Madejski (New York:
  AIP), 801, 345, \dodoi{10.1063/1.2141897}

\bibitem[{{Spitkovsky}(2008)}]{spitkovsky2008}
{Spitkovsky}, A. 2008, \apjl, 682, L5, \dodoi{10.1086/590248}

\bibitem[{{Totani} \& {Kitayama}(2000)}]{totani2000}
{Totani}, T., \& {Kitayama}, T. 2000, \apj, 545, 572, \dodoi{10.1086/317872}

\bibitem[{{van Weeren} {et~al.}(2019){van Weeren}, {de Gasperin}, {Akamatsu},
  {Br{\"u}ggen}, {Feretti}, {Kang}, {Stroe}, \& {Zandanel}}]{vanweeren2019}
{van Weeren}, R.~J., {de Gasperin}, F., {Akamatsu}, H., {et~al.} 2019, \ssr,
  215, 16, \dodoi{10.1007/s11214-019-0584-z}

\bibitem[{{van Weeren} {et~al.}(2011){van Weeren}, {Hoeft}, {R{\"o}ttgering},
  {Br{\"u}ggen}, {Intema}, \& {van Velzen}}]{vanweeren2011}
{van Weeren}, R.~J., {Hoeft}, M., {R{\"o}ttgering}, H.~J.~A., {et~al.} 2011,
  \aap, 528, A38, \dodoi{10.1051/0004-6361/201016185}

\bibitem[{{van Weeren} {et~al.}(2010){van Weeren}, {R{\"o}ttgering},
  {Br{\"u}ggen}, \& {Hoeft}}]{vanweeren2010}
{van Weeren}, R.~J., {R{\"o}ttgering}, H. J.~A., {Br{\"u}ggen}, M., \& {Hoeft},
  M. 2010, Science, 330, 347, \dodoi{10.1126/science.1194293}

\bibitem[{{Vazza} {et~al.}(2016){Vazza}, {Br{\"u}ggen}, {Wittor}, {Gheller},
  {Eckert}, \& {Stubbe}}]{vazza2016}
{Vazza}, F., {Br{\"u}ggen}, M., {Wittor}, D., {et~al.} 2016, \mnras, 459, 70,
  \dodoi{10.1093/mnras/stw584}

\bibitem[{{Vazza} {et~al.}(2009){Vazza}, {Brunetti}, \& {Gheller}}]{vazza2009}
{Vazza}, F., {Brunetti}, G., \& {Gheller}, C. 2009, \mnras, 395, 1333,
  \dodoi{10.1111/j.1365-2966.2009.14691.x}

\bibitem[{Weibel(1959)}]{Weibel1959}
Weibel, E.~S. 1959, Phys. Rev. Lett., 2, 83, \dodoi{10.1103/PhysRevLett.2.83}

\bibitem[{{Wittor} {et~al.}(2020){Wittor}, {Vazza}, {Ryu}, \&
  {Kang}}]{wittor2020}
{Wittor}, D., {Vazza}, F., {Ryu}, D., \& {Kang}, H. 2020, \mnras, 495, L112,
  \dodoi{10.1093/mnrasl/slaa066}

\end{thebibliography}
\bibliographystyle{aasjournal}

\end{document}